\documentclass[12pt]{article}
\pdfoutput=1
\usepackage{a4wide}
\usepackage{amssymb}
\usepackage{graphicx}
\usepackage{caption}
\usepackage{subcaption}
\usepackage{mathdots}

\def\drawbox#1#2{\hrule height#2pt 
        \hbox{\vrule width#2pt height#1pt \kern#1pt 
              \vrule width#2pt}
              \hrule height#2pt}

\def\Fund#1#2{\vcenter{\vbox{\drawbox{#1}{#2}}}}
\def\Asym#1#2{\vcenter{\vbox{\drawbox{#1}{#2}
              \kern-#2pt       % line up boxes
              \drawbox{#1}{#2}}}}

\def\funda{\Fund{6.5}{0.4}}
\def\asymm{\Asym{6.5}{0.4}}

\def\symm{\funda\kern-0.4pt\funda}

\def\makeatletter{\catcode`\@=11}% 11:letter
\makeatletter
\def\mathbox#1{\hbox{$\m@th#1$}}%
\def\math@ccstyles#1#2#3#4#5#6#7{{\leavevmode
      \setbox0\mathbox{#6#7}%
      \setbox2\mathbox{#4#5}%
      \dimen@ #3%
      \baselineskip\z@\lineskiplimit#1\lineskip\z@
      \vbox{\ialign{##\crcr
             \hfil \kern #2\box2 \hfil\crcr
             \noalign{\kern\dimen@}%
             \hfil\box0\hfil\crcr}}}}
\def\mathaccstyles{\math@ccstyles\maxdimen}
\def\maththroughstyles{\math@ccstyles{-\maxdimen}}
\def\unity%
 {\maththroughstyles{.45\ht0}\z@\displaystyle {\mathchar"006C}\displaystyle 1}
\def\be{\begin{eqnarray}}
\def\ee{\end{eqnarray}}

\begin{document}

\setcounter{table}{0}

\mbox{}
\vspace{2truecm}
\linespread{1.1}

%%%%%%%%%%%%%%%%%
\centerline{\LARGE \bf 5d quivers and their $AdS_6$ duals}
%Five-dimensional gauge theories and their $AdS$ duals } 

%\vspace{0.5cm}

%\centerline{\LARGE \bf }

\vspace{2truecm}

\centerline{
    {\large \bf Oren Bergman ${}^{a}$} \footnote{bergman@physics.technion.ac.il}
     {\bf and}
    {\large \bf Diego Rodr\'{\i}guez-G\'omez${}^{a}$} \footnote{drodrigu@physics.technion.ac.il}}

\vspace{1cm}
\centerline{{\it ${}^a$ Department of Physics, Technion, Israel Institute of Technology}} \centerline{{\it Haifa, 32000, Israel}}
\vspace{1cm}

\centerline{\bf ABSTRACT}
\vspace{1truecm}

\noindent
We consider an infinite class of 5d supersymmetric gauge theories involving products of symplectic and unitary groups
that arise from D4-branes at orbifold singularities in Type I' string theory.
The theories are argued to be dual to warped $AdS_6\times S^4/\mathbb{Z}_n$ backgrounds in massive Type IIA
supergravity. %extending the result of Brandhuber and Oz for the $n=1$ case.
In particular, this demonstrates the existence of supersymmetric 5d fixed points of quiver type.
We analyze the spectrum of gauge fields and charged states in the supergravity dual, and find
a precise agreement with the symmetries and charged operators in the quiver theories.
We also comment on other brane objects in the supergravity dual and their interpretation in the field theories.

\newpage

\tableofcontents
\section{Introduction}

Five dimensional gauge theories are non-renormalizable and therefore generically do not exist as microscopic theories.
In some cases, like the maximally supersymmetric Yang-Mills theory, it is believed that the UV theory is six-dimensional.
However there are a number of examples of strongly-coupled supersymmetric five-dimensional fixed point theories, corresponding to specific
gauge groups and matter content \cite{Seiberg:1996bd,Morrison:1996xf,Intriligator:1997pq}.

A key ingredient is that the minimal supersymmetry in five dimensions has eight supercharges, and therefore imposes strong restrictions. 
As in 4d $\mathcal{N}=2$ gauge theories, a supersymmetric 5d gauge theory is defined by a prepotential ${\cal F}$.
In 5d ${\cal F}$ can be at most cubic, because the 5d Chern-Simons coupling, defined by the third derivative of ${\cal F}$,
must be quantized to preserve gauge-invariance.
Furthermore, quantum effects are restricted to one-loop shifts of the cubic term 
\cite{Witten:1996qb}. 
The exact prepotential on the Coulomb branch 
for any gauge group and any matter hypermultiplet content is given by \cite{Intriligator:1997pq}
\be
\label{prepotential}
{\cal F} = \frac{1}{2g_0^2}h_{ij}\phi^i\phi^j
+ \frac{c_0}{6}d_{ijk}\phi^i\phi^j\phi^k
+ \frac{1}{12}\left(
\sum_{\bf{R}} |{\bf{R}}\cdot{\phi}|^3 -
\sum_f\sum_{{\bf {w}}\in {\bf W}_f} 
|{\bf{w}}\cdot{\phi} + m_f|^3\right) \,,
\ee
where $h_{ij}=\mbox{Tr}(T_i T_j)$, $d_{ijk}=\frac{1}{2}\mbox{Tr}(T_i(T_jT_k+T_kT_j))$,
${\bf R}$ are the roots of the gauge group $G$ and ${\bf W}_f$ are the weights of $G$ in the matter representation $f$.
Generically the theory does not make sense beyond $\phi\sim 1/g_0^2$, where
the effective gauge coupling, defined by the second derivative of ${\cal F}$, diverges.
This reflects the non-renormalizability of the field theory.
However, by suitably choosing the matter content for a given gauge group such singularities can be avoided,
suggesting that a strongly-coupled fixed point with $g_0\rightarrow \infty$ exists.
A few of these theories can be engineered using 5-brane webs in Type IIB string theory \cite{Aharony:1997ju,Aharony:1997bh}.
This construction makes apparent the structure of the moduli space and the spectrum of BPS particles on the Coulomb branch.

In view of the AdS/CFT correspondence  
one cannot help but wonder about the AdS supergravity backgrounds
dual to these fixed points.
Unlike in other dimensions, the superconformal algebra in five dimensions is unique,
it's bosonic part being $SO(5,2)\times SU(2)_R$. It has half the amount of supersymmetry
of the maximally supersymmetric theories in $d=3,4$ and $6$.
Correspondingly, supersymmetric $AdS_6$ backgrounds cannot be obtained by dimensional reduction
on simple spaces like $S^4$ or $\mathbb{C}P^2$. 
Following a proposal in \cite{Ferrara:1998gv}, the supergravity dual of a class of five-dimensional fixed points 
with gauge group $USp(2N)$ 
was found in \cite{Brandhuber:1999np}
by considering D4-branes 
in Type I' string theory.
The dual backgrounds are warped products of $AdS_6$ and half of an $S^4$ in massive Type IIA supergravity \cite{Romans:1985tz}.
These solutions can also be described in terms of $F(4)$ gauged supergravity \cite{Romans:1985tw}, 
which in turn can be obtained by dimensionally reducing massive Type IIA supergravity on the warped $S^4$ \cite{Cvetic:1999un}.
In fact this is the only known class of supersymmetric $AdS_6$ solutions.

A natural way to generalize the supergravity duals corresponding to branes in flat space is to look at branes in orbifolds.
This gives rise to quiver gauge theories, namely to product gauge groups and bi-fundamental matter fields. 
We will consider D4-branes in $\mathbb{C}^2/\mathbb{Z}_n$ orbifolds of Type I' string theory.
The corresponding supersymmetric quiver gauge theories involve products of $USp$ and $SU$ groups, and 
matter hypermultiplets transforming in bi-fundamental and antisymmetric tensor representations.
This is T-dual to the construction of 6d quiver theories using D5-branes in Type I orbifolds 
\cite{Douglas:1996sw,Dabholkar:1996zi,hep-th/9604129}.
At large $N$, the dual supergravity backgrounds will have a warped $AdS_6\times S^4/\mathbb{Z}_n$ geometry.
A similar orbifold construction using D3-branes and orientifold 7-planes was used to derive supergravity duals of 4d quiver theories
of the same type \cite{hep-th/9808175,hep-th/0006140}.

At this point one might be discouraged by the following argument 
that appears to rule out 5d supersymmetric fixed points of quiver type \cite{Intriligator:1997pq}.
If two gauge group factors are connected by a matter field, one can see in (\ref{prepotential}) that
a non-zero VEV for an adjoint scalar in one of the gauge groups 
contributes with a minus sign to the effective coupling of the other gauge group,
necessarily leading to a singularity somewhere in the moduli space.
However, at least in some cases, this singularity coincides with the appearance of new massless states,
as seen in the Type IIB 5-brane web construction, which may allow a continuation past infinite coupling.
In these cases the existence of the fixed point can be argued using S-duality \cite{Aharony:1997ju}.
Our $AdS_6$ backgrounds lend further support to the existence of these and other fixed point theories.

\medskip

The plan for the rest of the paper is as follows.
In section 2 we will review the $USp(2N)$ theory, its realizations in string theory and its supergravity dual.
In section 3 we will describe the supersymmetric quiver gauge theories that are obtained from $\mathbb{Z}_n$ orbifolds of the 
$USp(2N)$ theory, and
in section 4 we will discuss their supergravity duals.
In particular, we will analyze the spectrum of gauge fields and charges in the dual backgrounds,
and compare them with the global symmetries and mesonic and baryonic operators in the quiver gauge theories.
We will also discuss supergravity objects corresponding to instanton operators, cosmic strings, domain walls
and baryon vertices in the gauge theories.
We conclude in section 5, where we also raise some open questions and suggest a number of generalizations.

\section{The $USp(2N)$ theory}
\subsection{Field theory}

The simplest example of a 5d fixed point is an ${\cal N}=1$ $SU(2)$ gauge theory
with $N_f$ matter hypermultiplets in the fundamental representation \cite{Seiberg:1996bd}.
The Coulomb branch of the moduli space is $\mathbb{R}^+$, parametrized by the scalar field $\phi$ in the vector
multiplet of $U(1)\subset SU(2)$.
The effective gauge coupling in this case is
\be
\label{gauge_coupling}
\frac{1}{g_{eff}^2} = \frac{1}{g_0^2} + 16|\phi| - \sum_{i=1}^{N_f}|\phi - m_i| - \sum_{i=1}^{N_f}|\phi+m_i| \,,
\ee
where $m_i$ are the masses of the matter fields.
A necessary condition for a fixed point to exist is that this is positive everywhere on the moduli space.
This is satisfied only for $N_f\leq 7$.
There is no bare CS coupling in this theory, since the third order Casimir $d_{ijk}$ vanishes.
The one-loop contribution on the Coulomb branch (for $m_i=0$) gives
\be
\label{CS_coupling}
c = 2(8-N_f) \,.
\ee

Other than the $SU(2)_R$ associated to the two pseudoreal supercharges of the minimal 5d supersymmetry, the 5d gauge theory also has 
an $SO(2N_f)\times U(1)_I$ global symmetry, where
$SO(2N_f)$ is the flavor symmetry associated to the fundamental hypermultiplets,
and $U(1)_I$ is the topological symmetry associated to the conserved instanton number current
\be
j = *\mbox{Tr}(F\wedge F) \,.
\ee
Note that the 5d CS term 
couples this current to the gauge field.
The instanton therefore acquires a $U(1)$ gauge charge on the Coulomb branch. 
It was also argued that the global symmetry is enhanced to the exceptional group $E_{N_f+1}$ at the fixed point,
due to the instanton becoming massless at the origin of the Coulomb branch of the fixed point theory
\cite{Seiberg:1996bd}. 

The simplest generalization of this theory is a $USp(2N)$ gauge theory with a matter hypermultiplet $A$ in the %``traceless" 
antisymmetric representation.
For $SU(2)=USp(2)$ this is a singlet. More generally the antisymmetric representation of $USp(2N)$
reduces to a traceless part with $A^{ab}J_{ab}=0$, where $J$ is the invariant antisymmetric tensor, and the singlet trace.
The results (\ref{gauge_coupling}) and (\ref{CS_coupling}) are essentially unchanged,
except that the Coulomb branch contains $N$ copies of the $SU(2)$ Coulomb branch $\phi_1,\ldots,\phi_N$.
There is an additional Higgs branch corresponding to $A$, at a generic point of which
the gauge symmetry is broken to $SU(2)^N$.
Since the number of broken generators is $N(2N+1) - 3N = 2N(N-1)$,
the dimension of the Higgs branch is $N(2N-1)-1 - 2N(N-1) = N-1$. 
There is also an additional ``mesonic" $SU(2)_M$ global symmetry, under which $A$ transforms as a doublet.
The corresponding meson operator is just the trace $M = \mbox{Tr}[A] = A^{ab} J_{ab}$. 
There also seems to be a baryonic Pfaffian operator 
$\mbox{Pf}(A) = \epsilon^{a_1\cdots a_{2N}}  A_{a_1 a_2}\cdots A_{a_{2N-1} a_{2N}}$,
but this is actually related to the meson as $\mbox{Pf}(A) \propto M^N$.

\subsection{Type IIB brane construction}

\begin{figure}[h!]
\centering
\includegraphics[scale=.7]{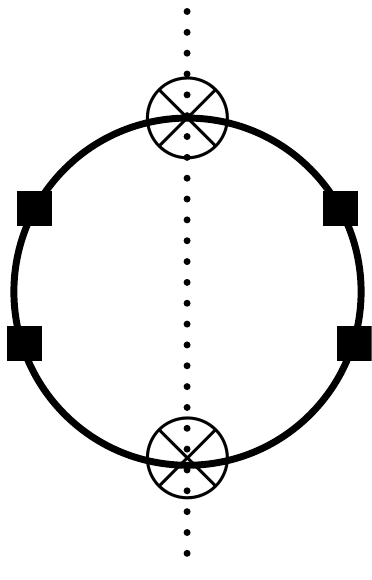} 
\caption{Type IIB brane configuration for the 5d $USp(2N)$ theory with flavors.
The crossed circles represent O7-planes and the squares are D7-branes.}
\label{CartoonNoOrb}
\end{figure}

The $USp(2N)$ theory can be realized by a brane configuration in Type IIB string theory 
on $\mathbb{R}^{1,4}\times \mathbb{R}^4\times S^1$,
with $2N$ D5-branes along $\mathbb{R}^{1,4}\times S^1$, two orientifold 7-planes along
$\mathbb{R}^{1,4}\times \mathbb{R}^3$ located at opposite points on the $S^1$,
and $N_f$ D7-branes parallel to the O$7$-planes together with their images, 
as follows (see also fig.~\ref{CartoonNoOrb}):
\begin{equation}
\label{braneconf} 
\begin{array}{l c |  c c c c c c c c c}
& 0 & 1 & 2 & 3 & 4 & \raisebox{.5pt}{\textcircled{\raisebox{-.9pt} {5}}} & 6 & 7 & 8 & 9 \\ \hline
 D7\,/\,O7^- & \times & \times & \times & \times & \times & & \times & \times & \times & \\ 
D5 & \times & \times & \times & \times & \times & \times & & & & \\ 
\end{array}
\end{equation}

The D5-branes come in pairs, as required by the consistency conditions of \cite{Gimon:1996rq}.
The orientifold maps the half of a D5-brane on one side of the circle to the half of its partner on the other side.
The flavors correspond to D5-D7 strings, and the antisymmetric hypermultiplet comes from
D5-D5 strings across an O7-plane.
The Coulomb branch corresponds to the positions of the D5-branes transverse to the O$7$-planes.
The part of the Higgs branch associated to the antisymmetric hypermultiplet corresponds
to the D5-brane positions along the O7-planes (together with the holonomy of the 
gauge field in the Cartan subalgebra along the $S^1$),
and the part associated to the fundamentals corresponds to breaking the D5-branes along the D7-branes.

\begin{figure}
\centering
       \begin{subfigure}[]{0.4\textwidth}
               \centering
               \includegraphics[width=\textwidth]{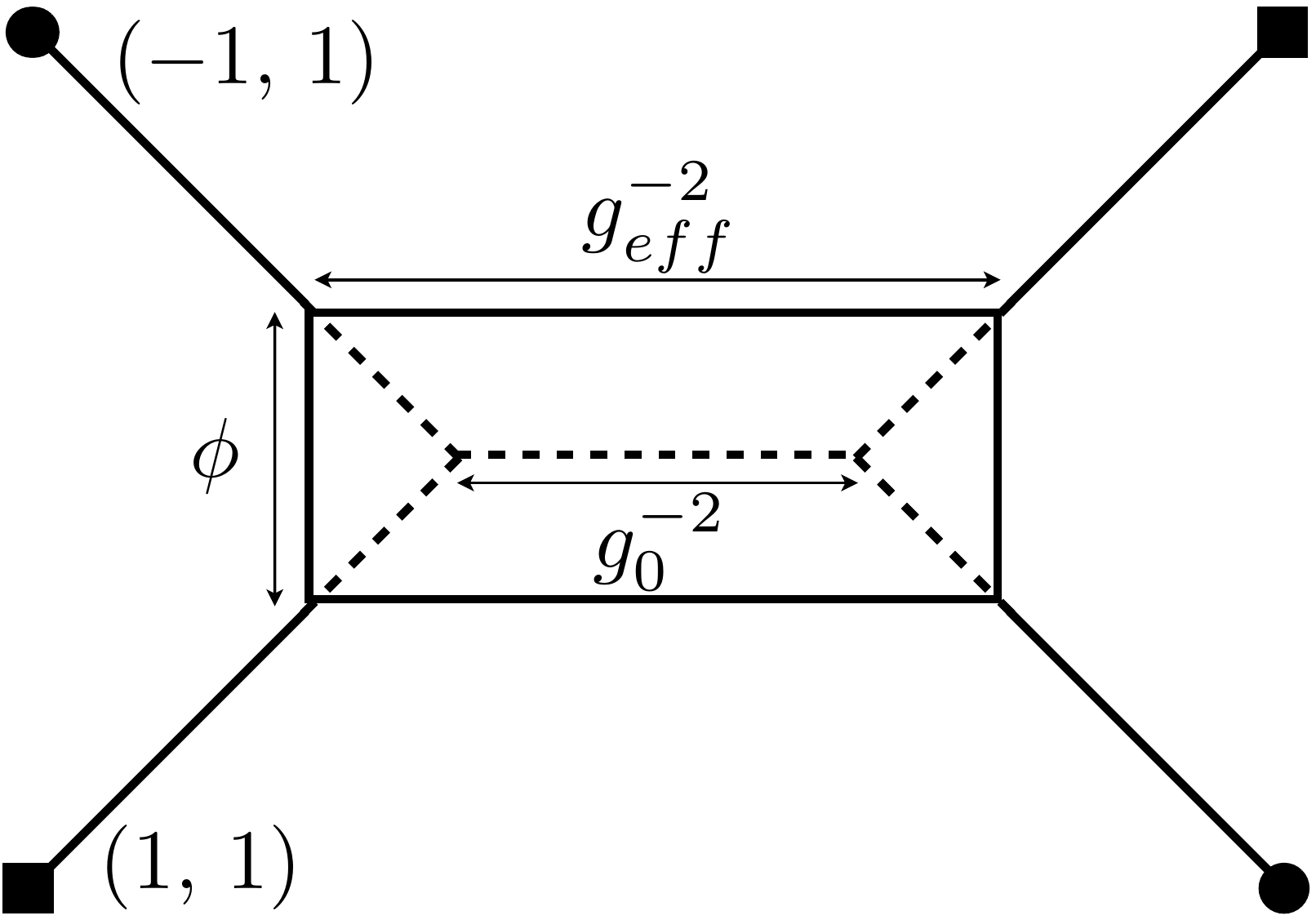}
               \caption{}
               \label{N=2OpenWebBOTHa}
       \end{subfigure}
       \hspace{1cm}
       \begin{subfigure}[]{0.29\textwidth}
               \centering
               \includegraphics[width=\textwidth]{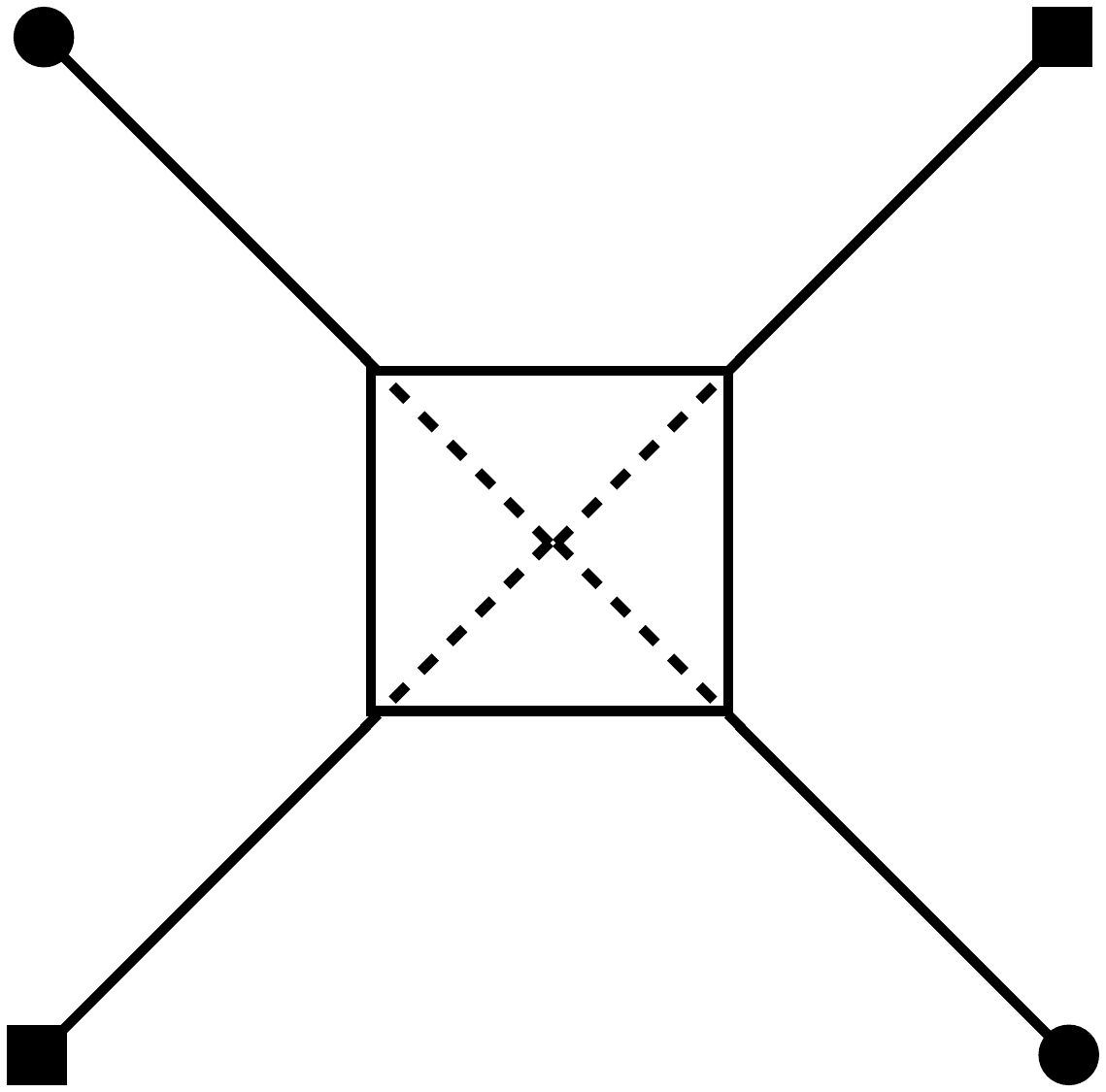}
               \caption{}
               \label{N=2OpenWebBOTHb}
       \end{subfigure}
             \caption{Pure $SU(2)$ web: (a) cutoff theory, (b) fixed point theory}
             \label{N=2OpenWebBOTH}
\end{figure}

This configuration provides a realization of the {\em classical} 5d theory.
In the quantum theory each O7-plane gets resolved into a pair of mutually non-local 7-branes, 
which in one representation are a $(1,1)$7-brane and a $(1,-1)$7-brane   \cite{Sen:1996vd, hep-th/9902179}.
The D5-branes then become a $(p,q)$5-brane web suspended between the four 7-branes in the $(x^5,x^9)$ plane.
For example for $SU(2)$ the web consists of two external $(1,1)$5-branes
and two external $(1,-1)$5-branes ending on the corresponding 7-branes,
and an internal rectangular face made of two parallel D5-branes and two parallel NS5-branes as shown in 
fig.~(\ref{N=2OpenWebBOTH}a) \cite{Aharony:1997ju}.\footnote{Naively two parallel D5-branes give a $U(2)$ 
gauge group, however the diagonal $U(1)$ is frozen 
by the 5d dynamics. The scalar in the $U(1)$ gauge multiplet corresponds
to a global web deformation that moves the 7-branes.} 
The length of the D5-brane segments corresponds to the inverse square effective coupling, and
the D5-brane separation to the Coulomb modulus $\phi$.
The bare coupling corresponds to the length of the D5-branes at zero separation. 
The fixed point theory is described by a web with a square face,
such that at the origin of the Coulomb branch only the external 5-branes remain, as shown in fig. (\ref{N=2OpenWebBOTH}b).

The spectrum of BPS states on the Coulomb branch is described by string webs supported by the 5-brane web \cite{Aharony:1997bh}.
In particular, a fundamental string suspended between the D5-branes corresponds to a $W$ boson with a mass 
proportional to $\phi$, and a D1-brane suspended between the NS5-branes corresponds to an instanton particle
with a mass proportional to $1/g_{eff}^2$.
Clearly all of these states are massless at the origin of the moduli space of the fixed point theory.

\begin{figure}[h!]
\centering
\includegraphics[scale=.4]{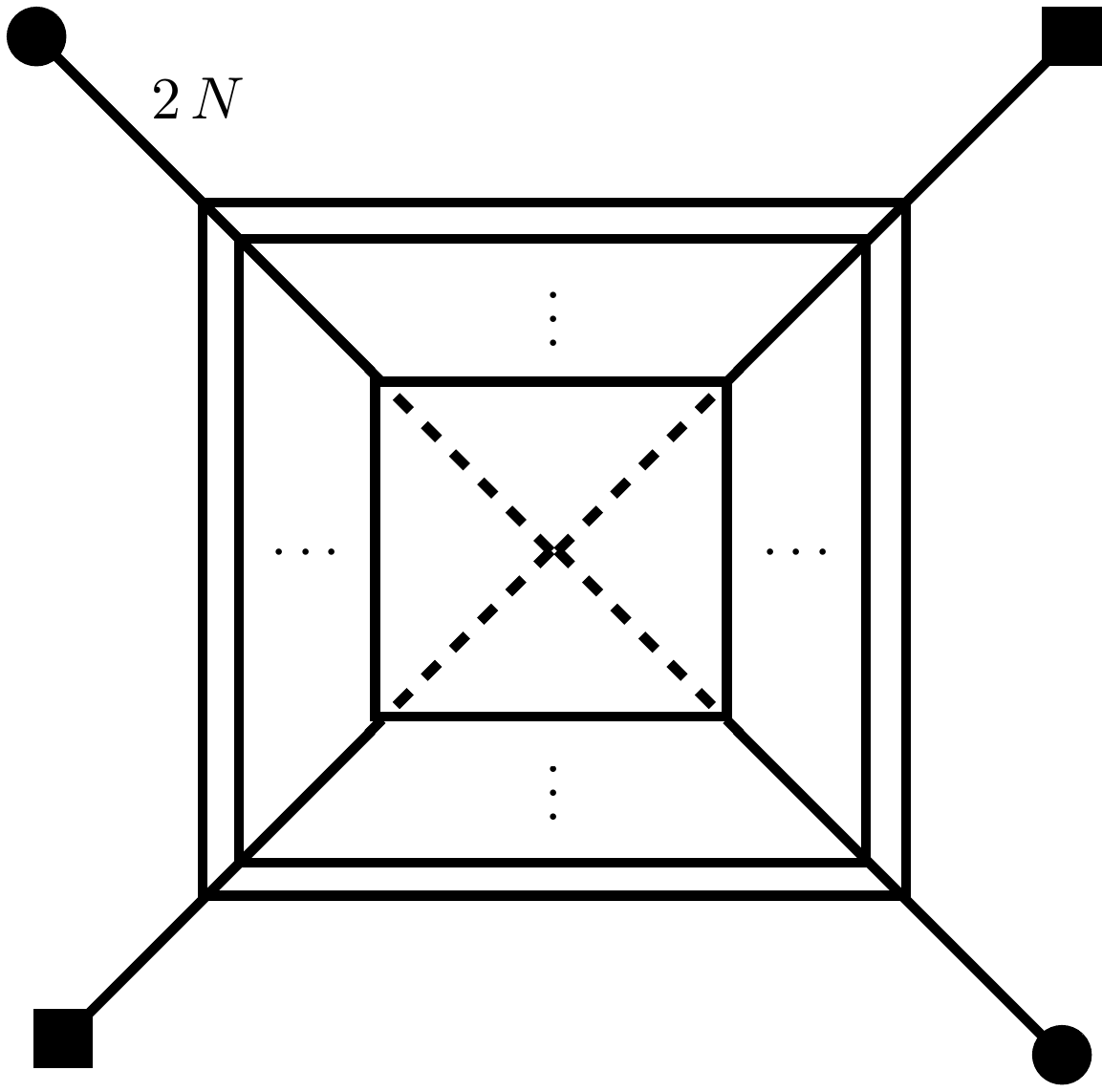} 
\caption{The $USp(2\,N)$ web.}
\label{GenericNOpenWeb}
\end{figure}

The 5-brane web for the $USp(2N)$
theory with an antisymmetric hypermultiplet is a simple generalization of the $SU(2)$ web,
consisting of $N$ copies of the latter (see Fig.~(\ref{GenericNOpenWeb})).
As usual, the moduli space corresponds to local web deformations.
In particular the Coulomb branch corresponds to the $N$ internal faces, in agreement with the field theory.
Note that in counting these we must apply the {\em generalized s-rule}, since all copies of a given external 5-brane 
end on the same 7-brane (see for example \cite{Benini:2009gi}).
The Higgs branch is described by the $N-1$ relative positions of the $N$ $SU(2)$ sub-webs
along the 7-branes in $(x^6,x^7,x^8)$, also in agreement with the field theory.

\subsection{Type I' brane construction}

While the Type IIB brane construction gives a nice geometrical description of the moduli space 
and BPS particle spectrum on the Coulomb branch,
it does not provide a good starting point for obtaining the supergravity duals.
For this purpose the Type I' brane construction is more useful.
T-duality relates the classical Type IIB configuration to a system of D4-branes in a background
with an O8-plane and $N_f$ D8-branes, which is the system originally used in \cite{Seiberg:1996bd}:\footnote{This is really part of a 
Type I' string theory background with two orientifold planes defining an interval and 16 D8-branes located 
at points along the interval. We are only concerned with the behavior of the D4-brane near one of the boundaries and so can 
ignore the other one.}
$$
\begin{array}{l c |  c c c c c c c c c}
& 0 & 1 & 2 & 3 & 4 & 5 & 6 & 7 & 8 & 9 \\ \hline
O8^-/D8 & \times & \times & \times & \times & \times & \times & \times & \times & \times & \\ 
D4 & \times & \times & \times & \times & \times & & & & & 
\end{array}
$$

Assume that the O8-plane is located at $x^9=0$,
and that the D8-branes are located at $x^9=x^9_i$, with $0\leq x^9_1\leq x^9_2 \leq \cdots \leq x^9_{N_f}$.
From here on we will take $\alpha'=1$.
The background is given by
\be
\label{O8D8_metric}
ds^2 &=& H_8^{-1/2}(x^9)\,dx_{1,8}^2 +H_8^{1/2}(x^9)\,(dx^9)^2 \\
\label{O8D8_dilaton}
e^{\Phi}&=&H_8^{-5/4}(x^9)\,,
\ee
with
\be
H_8(x^9) = a + 16 x^9 - \sum^{N_f} |x^9-x^9_i| - \sum^{N_f}|x^9+x^9_i|\,,
\ee
where $a$ is a constant.
There is also a piecewise constant RR 0-form field strength (``Romans mass"),
\be
\label{O8D8_RR}
F_0 = \frac{1}{4\pi}\left\{
\begin{array}{lc}
16 & 0<x^9<x^9_1 \\
16 - 2i & x^9_i < x^9 < x^9_{i+1} \,.
\end{array}
\right.
\ee
Therefore this is a background of massive Type IIA supergravity \cite{Romans:1985tz}.

The worldvolume gauge theory on the $N$ D4-branes plus their images is the 5d 
${\cal N}=1$ $USp(2N)$ theory with an 
antisymmetric hypermultiplet, and the D4-D8 strings give $N_f$ fundamental hypermultiplets.
The positions of the D4-branes in $x^9$ correspond to the Coulomb branch of the theory,
and their positions in $(x^5,x^6,x^7,x^8)$ to the part of the Higgs branch parameterized by the antisymmetric field.
The positions of the D8-branes in $x^9$ correspond to the flavor masses $m_i$.
Expanding the DBI action for a D4-brane in the background (\ref{O8D8_metric}), (\ref{O8D8_dilaton}) reproduces
the effective gauge coupling (\ref{gauge_coupling}).
The bare gauge coupling is identified with the constant $a$, and therefore the fixed point theory corresponds 
to setting $a=0$. 
In this case the dilaton blows up at $x^9=0$, and therefore the effective gauge coupling blows up
at the origin of the Coulomb branch.
The CS coupling (\ref{CS_coupling}) is derived from the RR worldvolume coupling 
$F_0 A\wedge F\wedge F$ when all $x^9_i=0$

We can understand the global symmetries of the gauge theory in this construction as follows.
The $SU(2)_R\times SU(2)_M$ part is realized as the $SO(4)$ rotation symmetry in $(x^5,x^6,x^7,x^8)$.
The flavor symmetry $SO(2N_f)$ corresponds to the 9d worldvolume gauge symmetry of the D8-branes
(when all $x^9_i=0$), and the instantonic $U(1)_I$ symmetry corresponds to the 10d RR 1-form potential.
Furthermore the stringy construction shows that, at the fixed point, the $SO(2N_f)\times U(1)_I$ part of the symmetry
is enhanced non-perturbatively to an exceptional group $E_{N_f+1}$, due to additional massless vectors described by
D0-branes localized at $x^9=0$ \cite{Polchinski:1995df,Matalliotakis:1997qe,Bergman:1997py}.
Note that D0-branes in the bulk of the 10d massive Type IIA background must have
semi-infinite strings attached to them \cite{Polchinski:1995sm}, and are therefore infinitely massive.
The states responsible for the gauge symmetry enhancement are special D0-brane states
that are localized at the 9d boundary. 
These either have no attached strings or only short strings \cite{Bergman:1997py}.

\subsection{Supergravity dual}

The Type I' construction can be used to obtain the large $N$ supergravity dual of the $USp(2N)$ fixed point CFT
\cite{Brandhuber:1999np}.
Consider the Type I' background with $N_f$ D8-branes on top of the O8-plane.
First, define a coordinate 
\be
\label{z}
z = \left(\frac{2x^9}{3}\sqrt{\frac{8-N_f}{2\pi}}\right)^{3/2} \,, 
\ee
in terms of which the O8-D8 background is conformally flat
\begin{equation}
\label{O8D8_background_2}
ds^2=\Omega^2(z)\left( dx_{1,8}^2 +dz^2\right)\,, \;\;\; e^{\Phi}= \Omega^5(z)\,, \;\;\;
F_0 = \frac{8-N_f}{2\pi \sqrt{\alpha'}} \,,
\end{equation}
where 
\begin{equation}
\Omega(z)=\Big(\frac{3}{4\pi}\,(8-N_f)\,z\Big)^{-1/6} \,.
\end{equation}
Note that the coordinate $z$ covers only the physical region on one side of the O8-plane.
The backreaction of the D4-branes introduces an additional warp factor and a 6-form flux:
\be
\label{backreacted_metric}
ds^2 &=& \Omega^2(z)\left[ H_4^{-1/2}(r)\,dx_{1,4}^2
+ H_4^{1/2}(r)\,\left(dx^2_{\mathbb{R}^4} + dz^2\right)\right] \\
\label{backreacted_dilaton}
e^{\Phi}&=&\Omega^{5}(z)\,H_4^{-1/4}(r)\\[5pt]
\label{backreacted_flux}
F_6 &=& d^5x \wedge dH^{-1}_4(r) \,,
\ee
where $r^2 = \tilde{r}^2+ z^2$ and $\tilde{r}^2=   x_5^2 + x_6^2 + x_7^2 + x_8^2 $.
In the near-horizon limit
\be
H_4(r) = \frac{Q_4}{r^{10/3}} \,,
\ee
where
\be
Q_4 = \left(\frac{2^{11} \pi^4}{3^4(8-N_f)}\right)^{1/3} N \,.
\ee
The precise relation between $Q_4$ and the number of D4-branes $N$
is obtained from the Gauss law
$\frac{1}{2\,\kappa_{10}^2}\,\int_{S^4} *F_6=N\,\mu_4$, where $\mu_p = 1/(2\pi)^p$ and
$\kappa_{10}=8\,\pi^{7/2}$.

Define an angular coordinate $\alpha$ by $\tilde{r}=r\cos\alpha$, $z=r\sin\alpha$. 
Expressed in terms of $\alpha$ and $u = r^{2/3}$, the background is seen to be a warped product of 
$AdS_6$ and $S^4$,
\begin{equation}
\label{AdS6}
ds^2= 
\hat{\Omega}^2(\alpha)
\left[ Q_4^{-1/2} u^2 \,dx_{1,\,4}^2
+\frac{9}{4} Q_4^{1/2} \frac{du^2}{u^2}
+ Q_4^{1/2} d\Omega_4^2 \right] \,,
\end{equation}
where
\begin{equation}
\hat{\Omega}(\alpha) = \left(\frac{3}{4\,\pi} (8-N_f)\sin\alpha\right)^{-1/6} \,,
\end{equation}
and
\be
\label{four_sphere}
d\Omega_4^2 = d\alpha^2+\frac{1}{4}\cos^2\alpha  
\left[\left(d\psi - \cos\theta d\phi\right)^2 + d\theta^2 + \sin^2\theta d\phi^2 \right] \,.
\ee
The coordinate ranges are given by $0\leq \theta \leq\pi$, $0\leq\phi\leq 2\pi$, $0\leq \psi \leq 4\pi$ and $0\leq \alpha\leq \pi/2$.
The internal space is therefore actually an $S^4$ hemisphere with an $S^3$ boundary at $\alpha=0$,
corresponding to the position of the O8-plane.
We can regard this as a full $S^4$ with $-\pi/2\leq \alpha \leq \pi/2$, modded out by the map $\alpha\rightarrow -\alpha$.
Strictly speaking this does
not follow from the flat space orientifold projection which takes $x_9\rightarrow -x_9$, 
as that would make $z$, and therefore
$\alpha$, imaginary. However it seems to be a consistent interpretation of the orientifold action in the near-horizon limit.
We will continue to denote the internal space as $S^4$, keeping in mind the orientifold action on $\alpha$.

The $AdS$ length scale is given by
\be
L = \frac{3}{2}\,Q_4^{1/4}=\frac{3^{2/3}\,\pi^{1/3}\,N^{1/4}}{2^{1/12}\,(8-N_f)^{1/12}}\, .
\ee
The warp factor, and therefore the curvature, diverges at $\alpha=0$. 
The dilaton is given by
\begin{equation}
e^{\Phi}=Q_4^{-1/4}\, \hat{\Omega}^5(\alpha)
\end{equation}
and it also diverges at the boundary $\alpha=0$.
For large $Q_4$ there is a region corresponding to $\sin\alpha \gg Q_4^{-3/10}$,
where both the dilaton and the curvature are small and the classical supergravity picture is valid.

The supergravity background has an $SO(2,5)$ symmetry corresponding to the isometry group 
of $AdS_6$, in agreement with the conformal symmetry of the 5d fixed point theory.
The gauge symmetry in the bulk includes the $SO(4) \sim SU(2)\times SU(2)$ subgroup
of the $SO(5)$ isometry group of $S^4$ preserved by the warping, in agreement with
the $SU(2)_R\times SU(2)_M$ part of the global symmetry of the field theory.
In particular,
the $U(1)$ part of the mesonic symmetry corresponds to shifts of the $\psi$ coordinate.
The (bosonic part of the) meson $M=\mbox{Tr}(A)$ is dual
to a $({\bf 2},{\bf 2})$ state. In particular this carries $1/2$ unit of KK momentum in $\psi$,
which is possible due to the $4\pi$ periodicity of $\psi$. 

The flavor $E_{N_f+1}$ symmetry is not visible in supergravity.
This is not surprising, since the flavor physics is localized at the boundary $\alpha=0$,
where the supergravity description breaks down.
Indeed, even the perturbative $SO(2N_f)$ flavor symmetry is not accessible due to the curvature singularity.
We should however be able to identify the 
instantonic symmetry $U(1)_I$ 
in the region where the supergravity description is valid.
This symmetry is dual to the bulk RR 1-form potential $C_1$. Correspondingly
instanton operators are dual to D0-branes.
As we mentioned previously, D0-branes in the bulk are accompanied by semi-infinite strings,
and therefore cannot correspond to local gauge-invariant operators.
Indeed, instantons are not gauge-invariant in this theory due to the CS coupling (\ref{CS_coupling}),
and must be accompanied by a semi-infinite Wilson line, in agreement with the bulk picture.

There are no additional particle-like wrapped D-brane states in this background, in agreement
with the absence of baryonic operators. This will change in the new models we consider below.

\section{5d orbifold theories}

We would like to generalize the above construction by replacing the $\mathbb{R}^4$ 
along $x^{5,\cdots, 8}$ by an ALE space asymptotic to $\mathbb{C}^2/\mathbb{Z}_n$.
The $\mathbb{Z}_n$ acts as $(z_1,z_2)\sim (e^{2\pi i/n} z_1, e^{-2\pi i/n} z_2)$, where
$z_1\equiv x^5 + ix^6$ and $z_2\equiv x^7 + ix^8$.

\subsection{Closed strings}
\label{ClosedStrings}

The background can be regarded as Type IIA string theory on the ALE space, then projected by $\Omega I_9$,
where $\Omega$ is worldsheet parity and $I_9$ is the reflection in $x^9$.
The metric on this ALE space is the Eguchi-Hanson metric
\be
\label{ALE_metric}
ds^2_{EH} = U d\vec{x}\,^2 + U^{-1}(d\varphi + \vec{A}\cdot d\vec{x})^2 \,,
\ee
where
\be
U(\vec{x}) = \sum_{i=1}^n \frac{1}{|\vec{x} - \vec{x}_i|} \,, \;\; \vec{\nabla}\times \vec{A} = - \vec{\nabla}U \,.
\ee
This ALE space has 2-cycles in $H_2(\mathbb{C}^2/\mathbb{Z}_n,\mathbb{Z}) = \mathbb{Z}^{n-1}$,
with a basis $\Sigma_i$ corresponding 
to the segments between $\vec{x}_i$ and $\vec{x}_{i+1}$. 
The ALE space admits three K\"ahler forms $\vec{\omega}$, whose periods are given by
\be
\int_{\Sigma_i} \vec{\omega} = \vec{x}_{i+1} - \vec{x}_i \equiv \vec{\zeta}_i \,,
\ee
and correspond to the $n-1$ blow-up parameters.
In the limit $\vec{x}_i\rightarrow 0$ the space degenerates to the orbifold $\mathbb{C}^2/\mathbb{Z}_n$,
and the 2-cycles shrink to zero size.

Type IIA string theory on $\mathbb{R}^{1,5}\times \mathbb{C}^2/\mathbb{Z}_n$
preserves $SO(1,5)\times SU(2)_R \times U(1)$ and ${\cal N}=(1,1)$ supersymmetry in six dimensions.
The untwisted sector includes a 6d $(1,1)$ gravity multiplet and a $(1,1)$ vector multiplet (Table~\ref{untwisted_sector}).
There are also $n-1$ twisted sectors associated to the 2-cycles $\Sigma_i$ of the ALE space, each of which
includes an additional $(1,1)$ vector multiplet (see Table~\ref{twisted_sector}).
In particular, the vector field originates from the reduction of the RR 3-form $C_3$ on $\Sigma_i$,
and the four scalars from the reduction of the NSNS 2-form $B_2$ on $\Sigma_i$
and from the blow-up modes $\vec{\zeta}_i$.

The orientifold projection reduces the spacetime symmetry to $SO(1,4)\times SU(2)_R\times U(1)$ and the supersymmetry to 5d ${\cal N}=1$.
The action of the orientifold combines worldsheet parity $\Omega$, which acts on the 10d massless fields as
\be 
\label{Omega_bulk}
\Omega: G_{MN}\rightarrow G_{MN} , \; \Phi \rightarrow \Phi , \; B_2 \rightarrow - B_2 , \; 
C_1\rightarrow C_1 , \; C_3\rightarrow -C_3 \,,
\ee
with the reflection $I_9$ that takes $x^9\rightarrow - x^9$.
The latter has the effect of exchanging the two $SU(2)$ factors in the 6d little group.
In the untwisted sector this leaves an ${\cal N}=1$ gravity multiplet, one vector multiplet and one hypermultiplet
(see Table~\ref{untwisted_sector}).
In the twisted sectors $\Omega$ exchanges the $j$th twisted sector (the sector twisted by $e^{2\pi ij/n}$)
with the $(n-j)$th twisted sector.
In the blown up ALE space this corresponds to exchanging the cycles $\Sigma_j \leftrightarrow \Sigma_{n-j}$.
For odd $n=2k+1$, all the twisted sectors are paired, leaving $k$ vector multiplets and $k$ hypermultiplets.
For even $n=2k$, $2k-2$ of the twisted sectors are paired, giving $k-1$ vector multiplets and hypermultiplets.
The $k$th twisted sector in the even orbifold is mapped to itself and must be treated separately.
As shown by Polchinski in the $\mathbb{Z}_2$ case \cite{hep-th/9606165}, 
there are two choices for the orientifold projection in this sector.
We can either project onto even or odd states (like discrete torsion in orbifolds).

In the first case we project out the RR $(3;1)$ state and the NSNS $(1;1)$ state,
since both $C_3$ and $B_2$ are odd under $\Omega$, leaving only the hypermultiplet.
This is called the ``orbifold without vector structure".
This is the ``ordinary" orbifold in the sense that the blow-up modes associated to the middle cycle $\Sigma_k$
remain, and the singularity can be fully resolved.
We can therefore continue to interpret the fields in the $k$th twisted sector as reductions of 10d fields on $\Sigma_k$.
An important subtlety in the orbifold without vector structure is that there is a discrete remnant $B$-flux
$b_k=\frac{1}{2\pi}\int_{\Sigma_k}B_2 = 1/2$.\footnote{Both $b_k= 0$ and $1/2$
are consistent with the orientifold projection due to the periodicity $b_k\sim b_k+1$,
but consistency conditions require $b_k=1/2$ \cite{hep-th/9703157}.
This is related to a non-trivial generalized second Stiefel-Whitney class \cite{hep-th/9605184}.}

In the second case we project out the RR $(1;1)$ state and the NSNS $(1;3)$ state, leaving only the vector multiplet.
This is the ``orbifold with vector structure".
The middle cycle is frozen at zero size in this case, but the modulus corresponding to the $B$-flux remains.
The perturbative orbifold point corresponds as usual to $b_i=\frac{1}{2}$ on all the cycles.

\begin{table}[h!]
\begin{center}
\begin{tabular}{|l|l|l|}
  \hline 
  \multicolumn{2}{|c|}{Type IIA $\mathbb{C}^2/\mathbb{Z}_n$} & 
  \multicolumn{1}{c|}{Type I' $\mathbb{C}^2/\mathbb{Z}_n$} \\\hline
  $G_{\mu\nu}$ & $(3,3;1) = (5;1)_+ +(3;1)_- +(1;1)_+$  & $(5;1)+(1;1)$\\
  $B_2$ & $(3,1;1)+(1,3;1) = (3;1)_+ + (3,1)_-$  & $(3;1)$ \\
  $\Phi$ & $(1,1;1) = (1;1)_+$   & $(1;1)$ \\
  $C_1$ & $(2,2;1) = (3;1)_+ + (1;1)_-$  & $(3;1)$ \\
  $C_3|_{\vec{\omega}}$ & $(2,2;3) = (3;3)_+ + (1;3)_+$ & $(1;3)$ \\ \hline
  $C_3$ & $(2,2;1) = (3;1)_+ + (1;1)_-$ & $(3,1)$ \\
   $B_2|_{\vec{\omega}}$ & $(1,1;3) = (1;3)_+$ & $\;\; -$ \\
   volume & $(1,1;1) = (1;1)_+$ & $(1;1)$ \\ \hline
\end{tabular}
 \end{center}
\caption{Untwisted sector: showing 6d $[SU(2)\times SU(2)]_{little}\times SU(2)_R$ charges
and 5d $SU(2)_{little}\times \mathbb{Z}_2 \times SU(2)_R$ charges, where $\mathbb{Z}_2$
is parity in the 6th coordinate.} 
\label{untwisted_sector}
\end{table}

\begin{table}[h!]
\begin{center}
\begin{tabular}{|c|c|l|}
  \hline \multicolumn{2}{|c|}{Type IIA $\mathbb{C}^2/\mathbb{Z}_n$} & \multicolumn{1}{c|}{Type I' $\mathbb{C}^2/\mathbb{Z}_n$} \\\hline
 $\int_{\Sigma_i} C_3$ & $(n-1)\left((3;1)_+ + (1;1)_-\right)$  &  
 $\begin{array}{ll}
 k(3;1) + k(1;1) & n=2k+1 \\
 (k-1)(3;1)+k(1;1) & n=2k ,\,\mbox{no VS}\\
 k(3;1)+(k-1)(1;1) & n=2k ,\,\mbox{VS}
 \end{array}$ \\\hline
 $\int_{\Sigma_i} B_2$ & $(n-1)(1;1)_+$ &
  $\begin{array}{ll}
  k(1;1) & n=2k+1 \\
 (k-1)(1;1) & n=2k,\,\mbox{no VS} \\
 k(1;1) & n=2k,\,\mbox{VS}
 \end{array}$ \\\hline
 blow-ups & $(n-1)(1;3)_+$ &
 $\begin{array}{ll}
 k(1;3) & n=2k+1  \\
 k(1;3) & n=2k,\,\mbox{no VS} \\
 (k-1)(1;3) & n=2k,\,\mbox{VS}
 \end{array}$ \\\hline
\end{tabular}
 \end{center}
 \caption{Twisted sectors: showing on the 5d $SU(2)_{little}\times \mathbb{Z}_2 \times SU(2)_R$ charges.}
 \label{twisted_sector}
\end{table}

\subsection{Quiver theories}

The worldvolume theories on D4-branes in the Type I' orbifold backgrounds are 5d quiver gauge theories.
They are essentially 5d versions of the 6d quiver theories on D5-branes in Type I orbifolds \cite{Douglas:1996sw}.
The results are summarized in Table~\ref{quiver_theories}.
Their structure is most easily understood using the T-dual Type IIB brane configurations, which
in addition to the O7-planes and D5-branes 
contain $n$ NS5-branes along $(x^0,x^1,x^2,x^3,x^4,x^9)$
and located, in a reflection symmetric way, at different positions on the circle.
For simplicity, we will omit from our discussion the fundamental matter fields corresponding to the D8-branes in the Type I' description
and to D7-branes in the Type IIB description.
These will be easy enough to incorporate in the supergravity duals.
Let us examine each case separately.
\begin{table}[h!]
\begin{center}
\begin{tabular}{|l|l|l|}
\hline 
model  & gauge group & matter \\ \hline 
$\mathbb{Z}_{2k}$ VS & $USp(2N)\times SU(2N)^{k-1}\times USp(2N)$   &  $\sum_{i=1}^k(\funda_i,\funda_{i+1})$  \\[5pt]
$\mathbb{Z}_{2k}$ no VS & $SU(2N)^k$   &  $\sum_{i=1}^{k-1}(\funda_i,\funda_{i+1}) + \asymm_1 + \asymm_{k}$  \\[5pt]
$\mathbb{Z}_{2k+1}$ & $USp(2N)\times SU(2N)^k$   &  $\sum_{i=1}^{k}(\funda_i,\funda_{i+1}) + \asymm_{k+1}$ \\
\hline
\end{tabular}
 \end{center}
\caption{5d orbifold quiver gauge theories} 
\label{quiver_theories}
\end{table}

\subsubsection{Even orbifolds with vector structure}

\begin{figure}[h!]
\centering
\includegraphics[scale=.5]{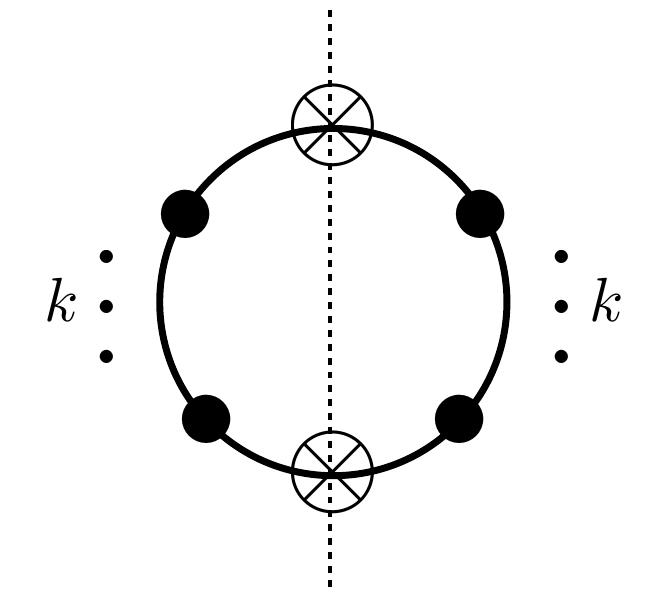}
\hspace{1cm} 
\raisebox{.8cm}{ \includegraphics[scale=.5]{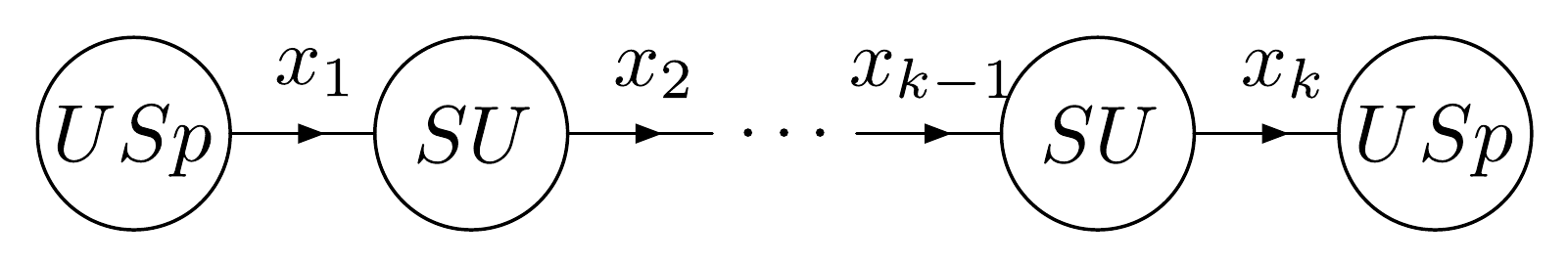} }
\caption{Type IIB configuration and quiver for even orbifolds with vector structure.}
\label{cartoonEVENVS}
\end{figure}

In this case there are $2k$ NS5-branes located symmetrically at arbitrary points on the two sides of the circle as in fig. (\ref{cartoonEVENVS}).
This divides the D5-branes into $2k$ segments, $2k-2$ of which are paired by the orientifold projection,
resulting in the product of gauge groups $USp(2N)\times SU(2N)^{k-1}\times USp(2N)$.
Classically, the gauge group on the paired segments is $U(2N)$, but the 5d dynamics freezes out the overall $U(1)$. 
The $k$ independent positions of the NS5-branes on the circle correspond to the $k$ relative gauge couplings.
(The sum of the gauge couplings corresponds to the size of the circle.)
In the Type I' orbifold background these correspond to the $k$ twisted NSNS $(1;1)$ modes.
The NS5-branes can also move pairwise in the $(x^6,\,x^7,\,x^8)$ directions,
their relative positions corresponding to the $k-1$ blow-up modes of the orbifold.
In the classical theory these correspond to the $k-1$ FI parameters associated to the $U(2N)$ factors.
There are also $k$ matter hypermultiplets $x_i$ in bi-fundamental representations of neighboring group factors,
that come from open strings between neighboring segments.

\begin{figure}
\centering
       \begin{subfigure}[]{0.3\textwidth}
               \centering
               \includegraphics[width=\textwidth]{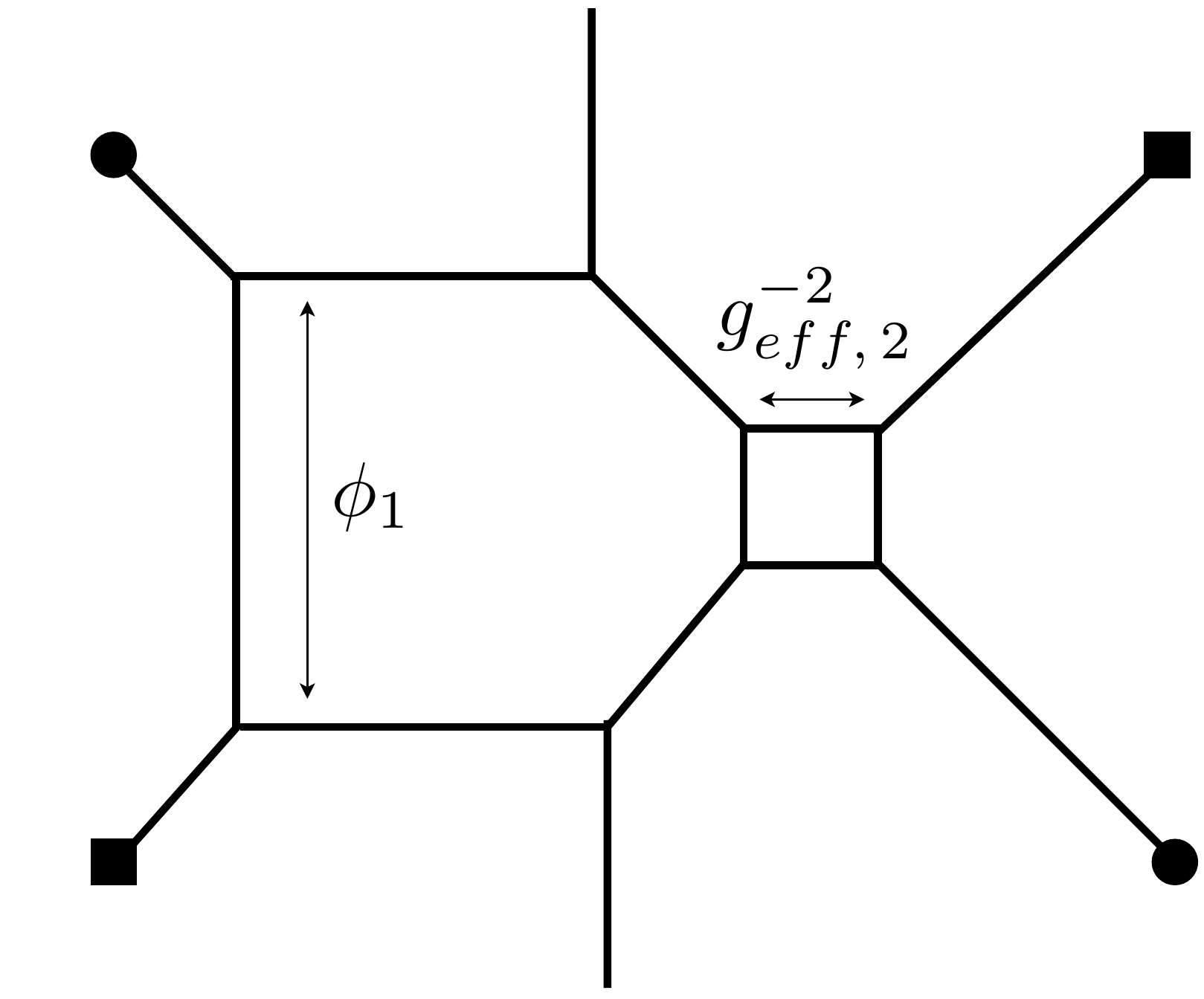}
               \caption{}
               \label{SU(2)xSU(2)a}
       \end{subfigure}
       \hspace{0.5cm}
       \begin{subfigure}[]{0.3\textwidth}
               \centering
               \includegraphics[width=\textwidth]{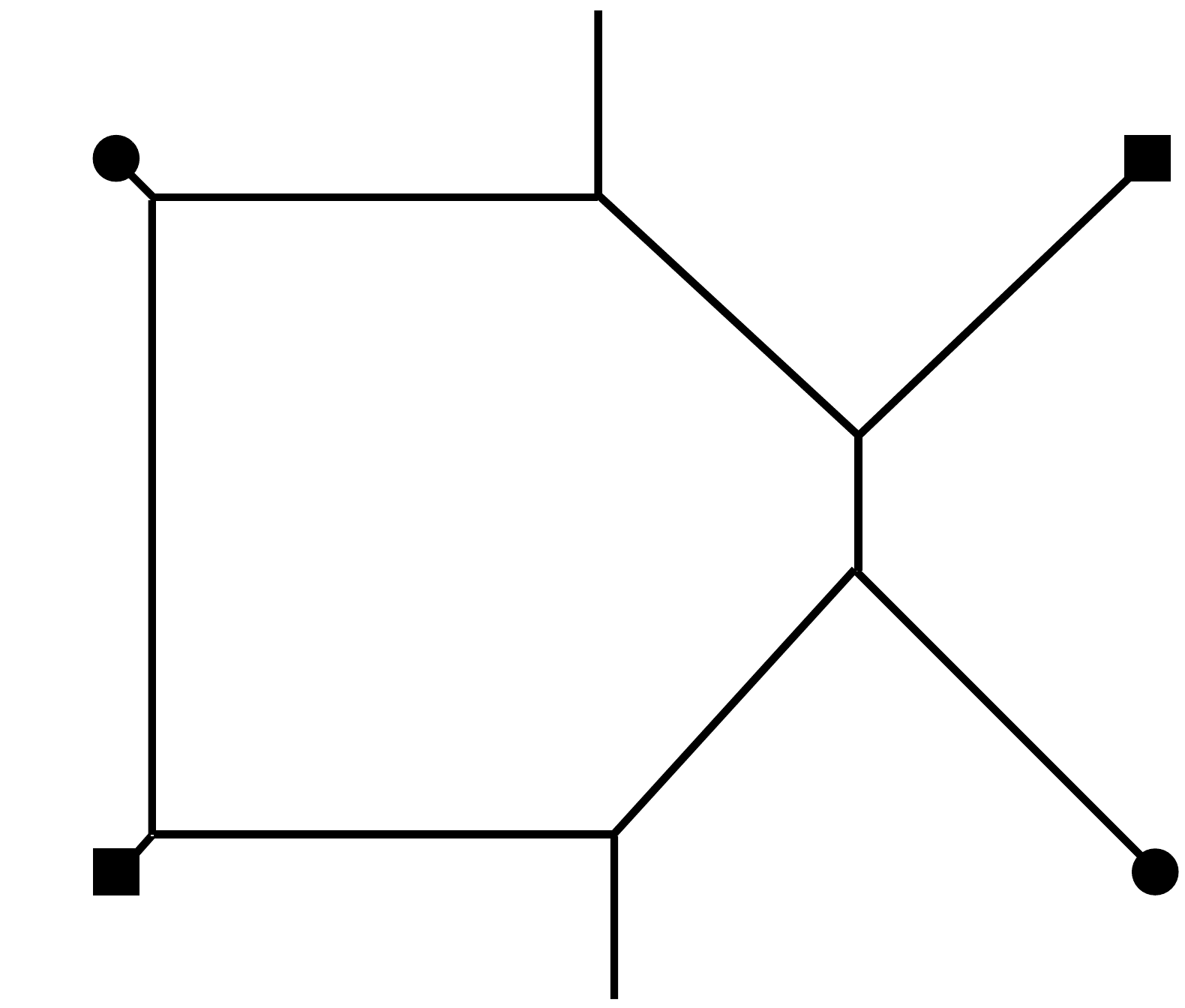}
               \caption{}
               \label{SU(2)xSU(2)b}
       \end{subfigure}
       \hspace{0.5cm}
        \begin{subfigure}[]{0.3\textwidth}
               \centering
               \includegraphics[width=\textwidth]{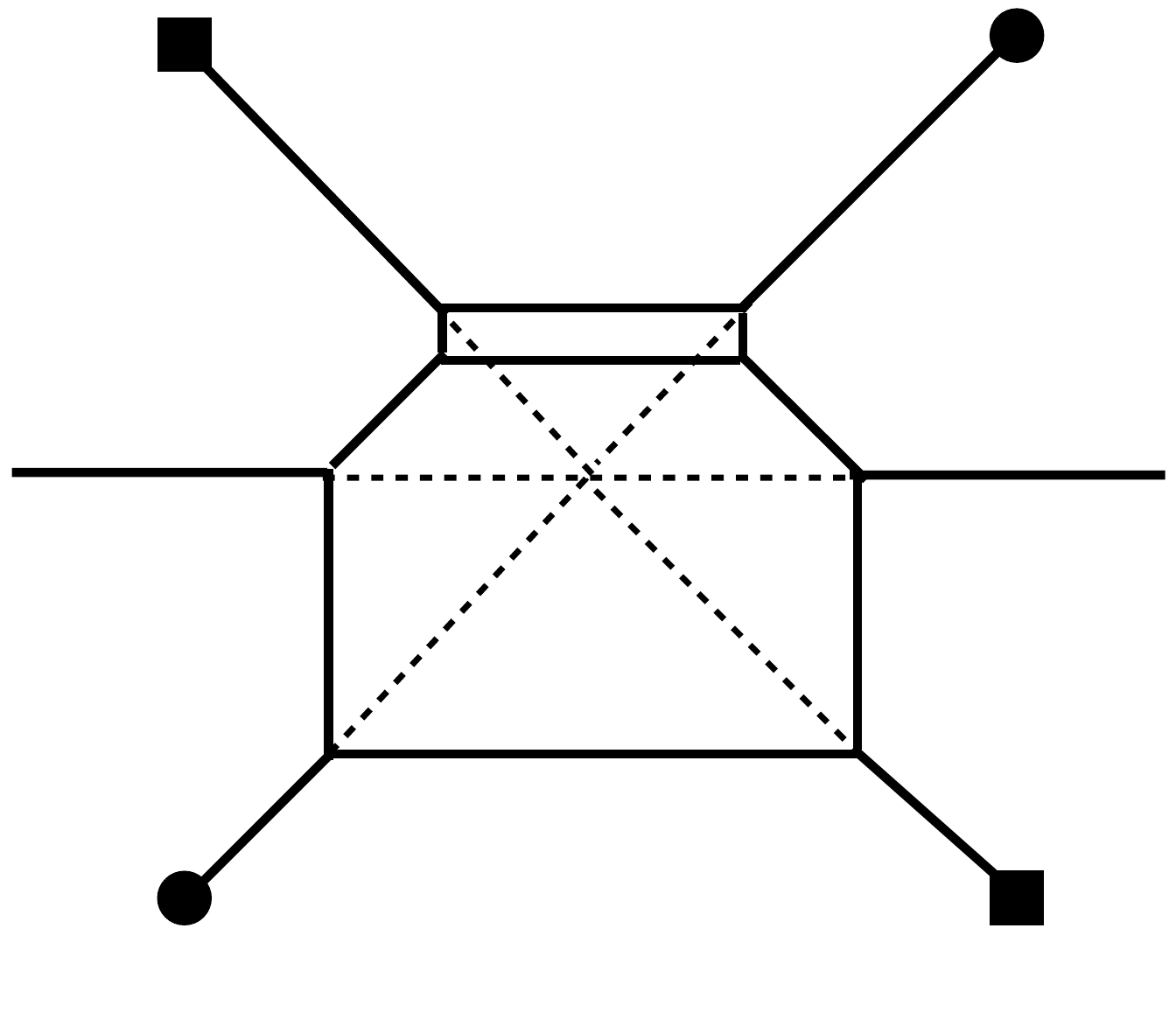}
               \caption{}
               \label{SU(2)xSU(2)b}
       \end{subfigure}
 \caption{Web for $SU(2)\times SU(2)$: (a) A generic point in the Coulomb branch, (b) the singularity in the Coulomb branch,
 (c) the S-dual web with $SU(3)$ and $N_f=2$. Since the latter can be collapsed, we expect the corresponding field theory to be a CFT.}
             \label{SU(2)xSU(2)}
\end{figure}

\begin{figure}[h!]
\centering
\includegraphics[scale=.4]{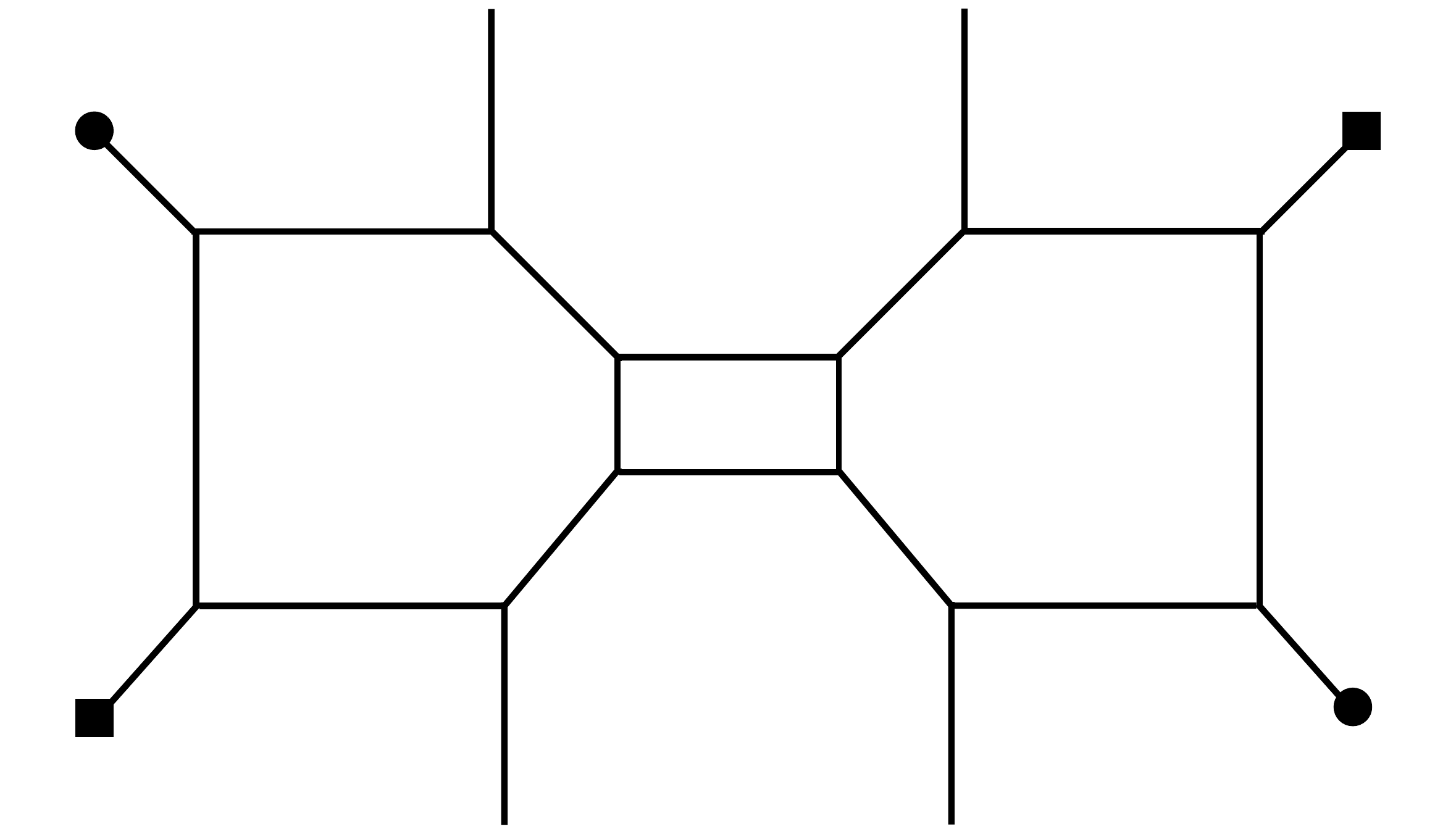}
\caption{Web corresponding to the $USp(2)\times SU(2) \times USp(2)$ theory.
The S-dual theory has $G=SU(4)$ and 4 fundamental hypermultiplets.}
\label{Z6GenericPoint}
\end{figure}

As in the parent $USp(2N)$ theory, the quantum dynamics of the quiver theory in the orbifold with vector structure
is captured by the resolution of the O7-planes into 7-branes.
This gives a 5-brane web with additional external NS5-branes.
For example, the 5-brane web for the theory with $k=1$ and $N=1$, namely $G=SU(2)\times SU(2)$ 
with a bi-fundamental,
is shown in fig.~\ref{SU(2)xSU(2)}a.\footnote{This again shows that the 
dynamics freeze out the $U(1)$ factors in the classical $U(2)\times U(2)$ gauge symmetry.}
The brane web exhibits clearly the singularity on the Coulomb branch:
as $\phi_1$ is increased, at some point $g_{eff,2}$ blows up (fig.~\ref{SU(2)xSU(2)}b).
The authors of \cite{Aharony:1997ju} proposed that at this point one should view the configuration
in an S-dual frame, namely rotated by 90 degrees, with NS5-branes and D5-branes exchanged (fig.~\ref{SU(2)xSU(2)}c).
This describes a 5d gauge theory with $G=SU(3)$ and two fundamental hypermultiplets, which 
has a well-defined strong coupling fixed point.
We can understand the singularity in the original quiver theory as due to an instanton particle 
corresponding to a D1-brane in the second face becoming massless.
In the S-dual picture this is simply a massless W-boson, which gives an enhanced gauge symmetry.
This argument can be generalized to the theories with vector structure for any $k$ and $N$.
The S-dual web is expected to give a gauge theory with $G=SU(2Nk-k+2)$ and $2k$ fundamentals \cite{WIP}.
(See, for example, the web for $k=2$ and $N=1$ 
in fig.~\ref{Z6GenericPoint}). 
All of these are expected to have well-defined fixed points according to \cite{Intriligator:1997pq}.

In preparation for a comparison with the proposed supergravity dual, let us consider the global
symmetries and the spectrum of gauge invariant operators charged under them.
First, there are $k+1$ instantonic $U(1)_I$ symmetries, one for each gauge group factor, and correspondingly
there are $k+1$ types of instantons.
As before, the instantons acquire gauge charges on the Coulomb branch due to CS interactions,
and must therefore be accompanied by semi-infinite Wilson lines.

Second, there is a $U(1)$ symmetry for each 
bi-fundamental $x_i$, acting as
$x_i\rightarrow e^{i\alpha} x_i$. 
Actually for $k=1$ it is enhanced to $SU(2)$, since the one bi-fundamental of $USp(2N)\times USp(2N)$ is pseudoreal.
(We will see this in the supergravity dual as well.)
A basis of gauge invariant operators charged under these symmetries can be obtained as follows.
First, we can form a meson using all the bi-fundamental fields: 
\be
M&=&\mbox{Tr}\Big[\prod_{i=1}^k x_i\Big]^2 \nonumber \\ 
&=& \Big[(x_1)^a_{\;\alpha_1} (x_2)^{\alpha_1}_{\;\alpha_2} \cdots 
(x_k)^{\alpha_{k-1}}_{\; b}\Big]
\Big[(x_1)^c_{\;\beta_1} (x_2)^{\beta_1}_{\;\beta_2} \cdots 
(x_k)^{\beta_{k-1}}_{\; d}\Big] \, J_{ac} J^{bd} \,,
\ee
where latin and greek indices are $USp$ and $SU$ indices, respectively.
We can also form $k$ di-baryons $B_i=\mbox{det} (x_i)$, or more explicitly
\be
B_i &=& \epsilon_{\alpha_1 \cdots \alpha_{2N}}\epsilon^{\beta_1\cdots \beta_{2N}}
(x_i)^{\alpha_1}_{\;\;\beta_1}\cdots (x_i)^{\alpha_{2N}}_{\;\;\beta_{2N}} \;\;\;\;\;\;\;\;\;\; i=2,\ldots,k-1 \nonumber \\
B_1 &=& \epsilon_{a_1 \cdots a_{2N}} \epsilon^{\beta_1\cdots \beta_{2N}} 
(x_1)^{a_1}_{\;\;\beta_1}\cdots (x_1)^{a_{2N}}_{\;\;\beta_{2N}} \\
B_k &=& \epsilon_{\alpha_1 \cdots \alpha_{2N}}\epsilon^{{b}_1\cdots {b}_{2N}}
(x_k)^{\alpha_1}_{\;\;{b}_1}\cdots (x_k)^{\alpha_{2N}}_{\;\;{b}_{2N}} \nonumber \,.
\ee
Since the baryons and meson satisfy
$\prod_{i=1}^k B_i \propto M^N$,
we can choose a basis $\{M, B_1,\ldots B_{k-1}\}$.
Correspondingly, we will define the mesonic and baryonic charges as 
\be
Q_M &=& \frac{1}{2}\sum_{i=1}^k Q_i \\
Q_{B,i} &=& Q_i - Q_k \,.
\ee
The normalization of the mesonic charge is fixed by the $k=1$ case, where the mesonic symmetry 
is enhanced to $SU(2)_M$.
The charges and scaling dimensions of the matter operators are shown in Table~\ref{vector_structure_charges}.
\begin{table}[h!]
\begin{center}
\begin{tabular}{|c|c|c|c|}
  \hline 
   operator  & $Q_M$ & $Q_{B,j}$ & $\Delta$ \\\hline
   $x_i$ & $1/2$ & $\delta_{ij}$ & $3/2$ \\
  $M$ & $k$ & 0 & $3k$ \\
  $B_i$ & $N$ & $2N \delta_{i,j}$ & $3N$ \\
  \hline
\end{tabular}
 \end{center}
\caption{Mesons and baryons for $\mathbb{Z}_{2k}$ with vector structure.} 
\label{vector_structure_charges}
\end{table}

\subsubsection{Even orbifolds without vector structure}

\begin{figure}[h!]
\centering
\includegraphics[scale=.5]{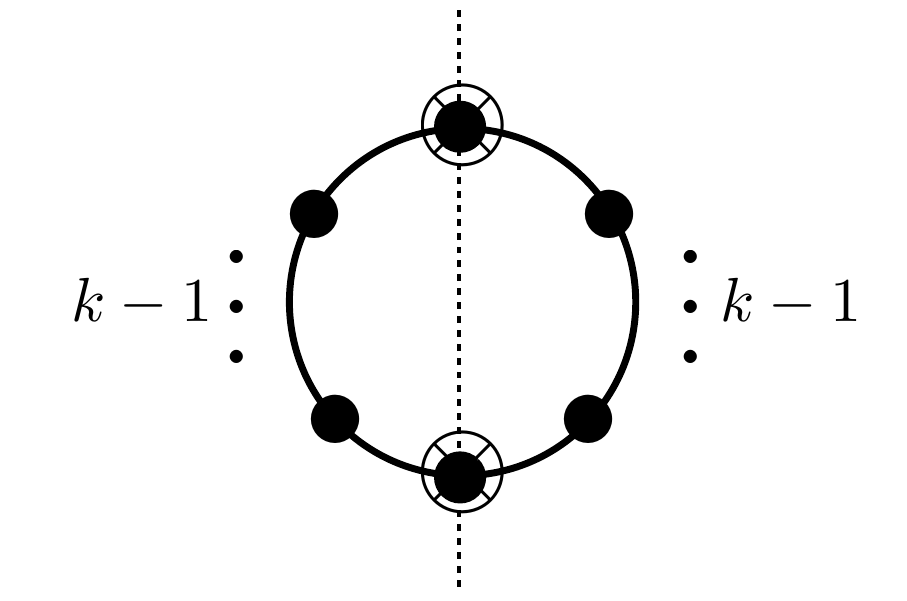}
\hspace{1cm} 
\raisebox{.8cm}{ \includegraphics[scale=.5]{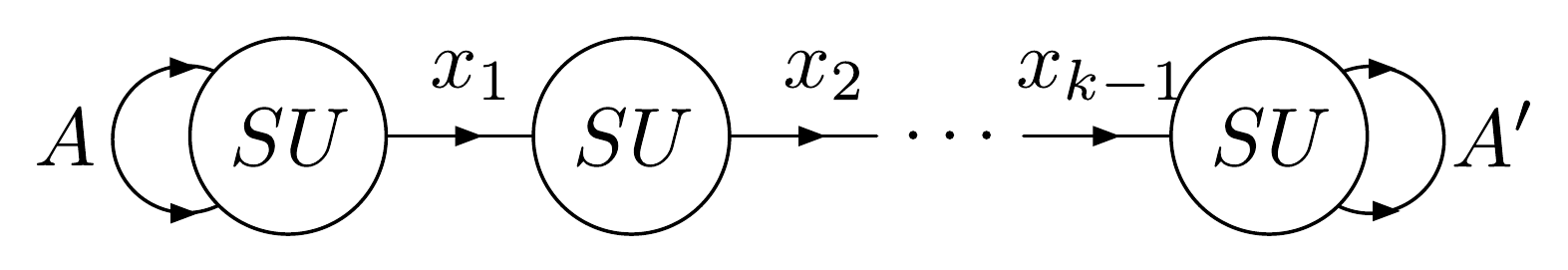} }
\caption{Type IIB configuration and quiver for even orbifolds without vector structure.}
\label{cartoonEVENNVS}
\end{figure}

The quiver theory for the orbifold without vector structure is constructed by placing $2k-2$ NS5-branes symmetrically 
at arbitrary points on the circle, and placing the remaining two at the location of each of the O7-planes as in fig.~\ref{cartoonEVENNVS}.
These two are ``stuck" in the circle direction.
Now all the segments are paired so there are $k$ $SU(2N)$ factors.
Correspondingly, there are $k-1$ relative gauge couplings described by the $k-1$ positions of the ``free" NS5-branes 
on the circle, and $k$ (classical) FI parameters described by the $k-1$ independent positions of the ``free" NS5-branes
along $(x^6,x^7,x^8)$ plus those of the two stuck NS5-branes (that can move independently), minus the center of mass position.
There are also $k-1$ bi-fundamental hypermultiplets $x_i$ coming from open strings that straddle the ``free" NS5-branes,
and two antisymmetric hypermultiplets $A,A'$ of the first and last $SU(2N)$ factors, respectively, coming from open strings
that straddle the ``stuck" NS5-branes.

Unfortunately, we are not able to resolve the classical brane configuration into a 5-brane web as in the previous model,
since we do not know how to resolve an O7-plane with a stuck NS5-brane.
Therefore we cannot use the S-duality argument.
However we will continue to assume that the dynamics freeze out the $U(1)$ parts of the $U(2N)$ gauge groups.

The global symmetries include $k$ instantonic $U(1)_I$'s and $k+1$ matter $U(1)$'s associated to the bi-fundamentals 
$x_i$ and the antisymmetrics $A,A'$, except for the case $k=1$, where
the matter symmetry associated to $A,A'$ is enhanced to $U(2)$.
The basis of charged gauge-invariant operators can be chosen as $\{M,B_A,B_1,\ldots, B_{k-1}\}$, where the meson is given by
\be
M =  \mbox{Tr}\Big[A\prod_{i=1}^{k-1}x_i^2 A' \Big]    \,,
\ee
$B_i$ are the di-baryons
\be
B_i =  \mbox{det}(x_i) \;\;\;\; i=1,\ldots, k-1\,,
\ee
and $B_A$ is a Pfaffian baryon operator associated to the antisymmetric field $A$.
\be
B_A=\mbox{Pf}(A) = \epsilon^{\alpha_1\cdots \alpha_{2N}} A_{\alpha_1 \alpha_2}\cdots A_{\alpha_{2N-1}\alpha_{2N}} \,.
\ee
Unlike in the $USp(2N)$ theory, the Pfaffian state is non-trivial here, since it is made of $SU(2N)$ antisymmetrics.
The additional baryon $B_{A'} = \mbox{Pf}(A')$ is related to these by 
$B_A \, B_{A'} \prod_{i=1}^{k-1} B_i  \propto M^N$.
We define the mesonic and baryonic charges in this case as 
\be
Q_M &=& \frac{1}{2}(Q_A + Q_{A'}) + \frac{1}{2}\sum_{i=1}^{k-1} Q_i \\
Q_{B,i} &=& Q_i - Q_A - Q_{A'} \\
Q_{B,A} &=& Q_A - Q_{A'} \,.
\ee
As before, the normalization of the mesonic charge is fixed by the $k=1$ case, where the symmetry is $SU(2)_M$.
The charges and scaling dimensions of the different operators are shown in Table~\ref{no_vector_structure_charges}.
\begin{table}[h!]
\begin{center}
\begin{tabular}{|c|c|c|c|c|}
  \hline 
   operator  & $Q_M$ & $Q_{B,j}$ & $Q_{B,A}$ & $\Delta$ \\\hline
  $x_i$ & $1/2$ & $\delta_{ij}$ &0&$3/2$\\
  $A$ & $1/2$ & $-1$ & $1$ & $3/2$\\
  $A'$ & $1/2$ & $-1$ & $-1$ & $3/2$\\
  $M$ & $k$ & 0 & 0 & $3k$  \\
  $B_A$ & 0 & $-N$ & $N$ & $3N/2$ \\
  $B_i$ & $N$ & $2N\delta_{ij}$ & 0 & $3N$ \\
\hline
\end{tabular}
 \end{center}
\caption{Mesons and baryons for $\mathbb{Z}_{2k}$ without vector structure.} 
\label{no_vector_structure_charges}
\end{table}

\subsubsection{Odd orbifolds}

\begin{figure}[h!]
\centering
\includegraphics[scale=.5]{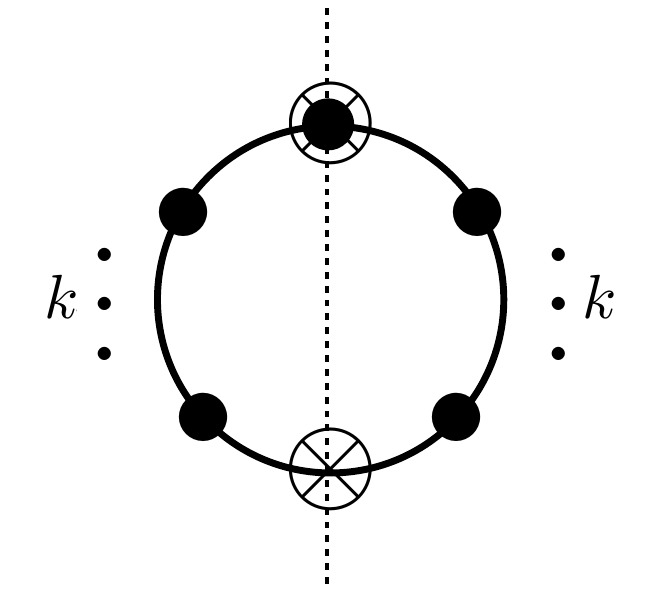} 
\hspace{1cm} 
\raisebox{.8cm}{\includegraphics[scale=.5]{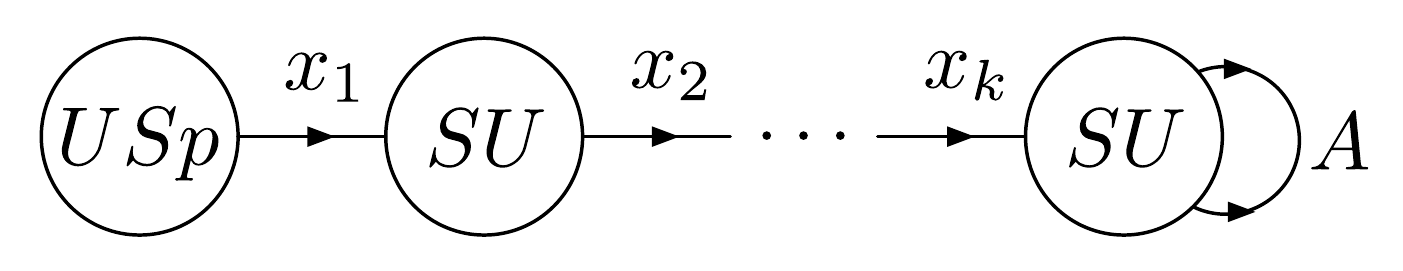} }
\caption{Type IIB configuration and quiver for odd orbifolds.}
\label{cartoonODD}
\end{figure}

The construction of the odd orbifold quivers is similar to the previous case in that it requires placing one NS5-brane 
on one of the O7-planes. The $2k$ remaining NS5-branes are distributed symmetrically on the circle as in fig. (\ref{cartoonODD}).
There is one unpaired segment that gives $USp(2N)$ and $2k$ paired segments that give $SU(2N)^k$.
The $k$ relative gauge couplings are seen as the $k$ positions of the free NS5-branes on the circle, and
the $k$ FI parameters 
correspond to the relative $(x^6,x^7,x^8)$ positions of the $k$ pairs of free NS5-branes and the stuck NS5-brane.
The matter content in this theory includes $k$ bi-fundamentals $x_i$ of neighboring group factors
and an antisymmetric $A$ of the last $SU(2N)$.
As in the previous model, we do not know how quantum effects resolve this configuration.

The global symmetries include $k+1$ instantonic $U(1)_I$'s and $k+1$ matter $U(1)$'s: $U(1)_A$ and $U(1)_i$ with $i=1,\ldots,k$.
We can choose the basis of gauge invariant charged operators as $\{M,B_1,\ldots,B_{k}\}$, where now
\be
M &=& \mbox{Tr}\Big[\prod_{i=1}^k x_i^2 A\Big] \nonumber\\
&=& \Big[(x_1)^a_{\;\alpha_1} (x_2)^{\alpha_1}_{\;\alpha_2}\cdots (x_k)^{\alpha_{k-1}}_{\;\alpha_k}\Big]
\Big[(x_1)^b_{\;\beta_1} (x_2)^{\beta_1}_{\;\beta_2}\cdots (x_k)^{\beta_{k-1}}_{\;\beta_k}\Big]
A^{\alpha_k \beta_k} J_{ab} \,,
\ee
and $B_i$ ($i=1,\ldots, k$), as usual, are the di-baryons.
The Pfaffian baryon $B_A=\mbox{Pf}(A)$ is related to these by
$B_A \prod_{i=1}^{k} B_i \propto M^N$.
We therefore define the mesonic and baryonic charges as 
\be
Q_M&=& \frac{1}{2}Q_A + \frac{1}{2}\sum_{i=1}^{k} Q_i\\
Q_{B,i} &=& Q_i - 2Q_A \,.
\ee
Although there is no enhancement of the mesonic symmetry in the odd orbifolds,
we include the $1/2$ in the normalization to be consistent with the $k=1$ cases of the even orbifolds.
The charge assignments and scaling dimensions are shown in Table~\ref{odd_orbifold_charges}.
\begin{table}[h!]
\begin{center}
\begin{tabular}{|c|c|c|c|}
  \hline 
   operator  & $U(1)_M$ & $U(1)_{B,j}$  & $\Delta$ \\\hline
   $x_i$ & $1/2$ & $\delta_{ij}$ & $3/2$ \\
   $A$ & $1/2$ & $-2$ & $3/2$ \\
  $M$ & $k+1/2$ & 0  & $3k+ 3/2$  \\
  $B_i$ & $N$ & $2N\delta_{ij}$ & $3N$ \\
\hline
\end{tabular}
 \end{center}
\caption{Mesons and baryons for $\mathbb{Z}_{2k+1}$.} 
\label{odd_orbifold_charges}
\end{table}

\subsubsection{Additional comments}

\begin{itemize}
\item We have only discussed quiver theories with equal rank gauge group factors.
In principle we could consider unequal ranks,
as well as odd-rank $SU$ group factors.
For example, the most general theory on the $\mathbb{Z}_{2k+1}$ orbifold has $G=USp(2N_0)\times \prod_{i=1}^{k} SU(N_i)$.
This corresponds to having fractional D4-branes of various types in the orbifold.
In the Type IIB construction different ranks correspond to different numbers of D5-branes in the different segments,
and an odd rank $SU(N)$ group is possible since the associated D5-branes do not cross the O7-plane.
By contrast, in 6d the chiral anomaly imposes strong constraints on the relative ranks of the gauge group factors   
\cite{Seiberg:1996qx, Danielsson:1997kt} (see also \cite{ hep-th/9702038, hep-th/9705030, hep-th/9712145}),
which can also be seen in the analogous Type IIA brane construction as a tadpole-cancellation condition on the NS5-branes \cite{hep-th/9712145}.
In 5d there are no continuous anomalies, and correspondingly no tadpoles in the brane configuration.
On the other hand, with a different number of D5-branes on each side, the NS5-brane becomes a $(1,q)$5-brane
(where $q$ is the difference in the number of D5-branes) and is bent by the appropriate angle.
This presumably leads to a constraint on the relative ranks.
\item Another possible generalization is to add bare CS terms.
This is possible only for the $SU(2N)$ factors (for $N>1$).
Related to this possibility, and the previous one, is the existence of domain walls separating
theories with different ranks and different CS levels, which we will comment on in the next section.
\item The Type IIB brane constructions (of the classical theories) can also be used to demonstrate
transitions between the different quiver theories.
For example, starting with a configuration corresponding to a $\mathbb{Z}_{2k}$ orbifold with vector structure,
bringing together a pair of NS5-branes on top of one of the O7-planes and then moving one of them far away along $(x^6,x^7,x^8)$, 
we end up with a configuration corresponding to a $\mathbb{Z}_{2k-1}$ orbifold. 
Repeating this process at the other O7-plane then leads to a $\mathbb{Z}_{2k-2}$ orbifold without vector structure.
These types of transitions where discussed in the 6d context using the analogous Type IIA brane configurations in \cite{hep-th/9712145}.
\end{itemize}

\section{Supergravity duals}

\subsection{Geometry}

In the near-horizon background the orbifold acts only on the internal space, so the geometry
is a warped product of $AdS_6$ and $S^4/\mathbb{Z}_n$.
The orbiold acts freely on the $S^3$ fiber giving the Lens space $S^3/\mathbb{Z}_n$.
The metric then has the same form 
as the one dual to the parent $USp(2N)$ theory (\ref{AdS6}), (\ref{four_sphere}), with the periodicity
of $\psi$ now $\psi\sim \psi + 4\pi/n$. 
Correspondingly, there is an additional factor of $n$ in the relation between $L$ and $N$ 
(incorporating also the $N_f$ D8-branes):
\be
L = \frac{3}{2}\,Q_4^{1/4}
= \frac{3^{2/3}\,\pi^{1/3}\,(nN)^{1/4}}{2^{1/12}\,(8-N_f)^{1/12}} \,.
\ee

\begin{figure}[h!]
\centering
\includegraphics[scale=.4]{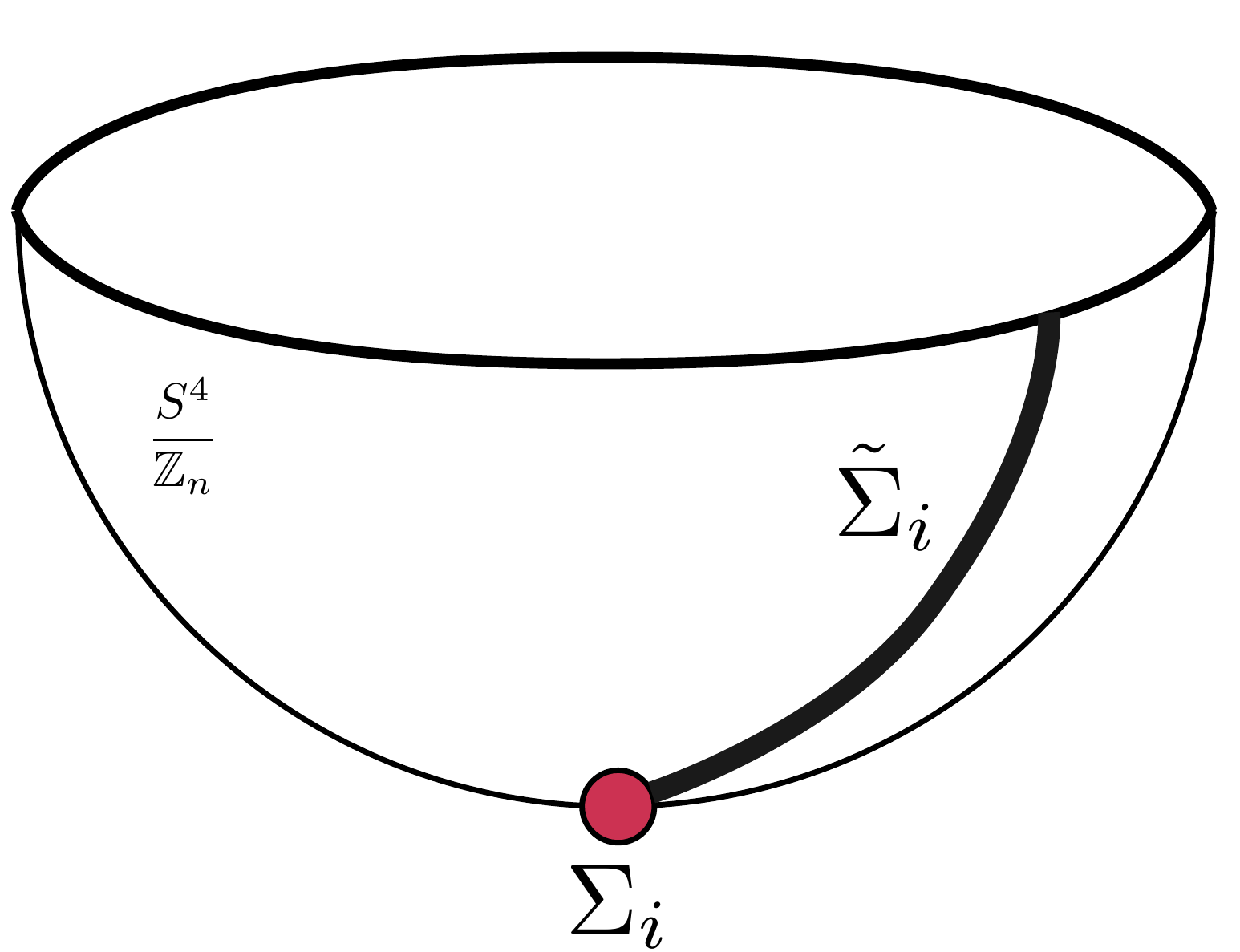}
\caption{Cartoon of the internal space.}
\label{CyclesCartoon}
\end{figure}

The space $S^4/\mathbb{Z}_n$ (again, we really mean an $S^4$ hemisphere)
has 
a fixed point singularity at $\alpha=\pi/2$,
around which it is locally $\mathbb{C}^2/\mathbb{Z}_n$. 
We must therefore include the twisted sectors in the near horizon background as well.
Using the geometric description,
there are $n-1$ vanishing 2-cycles $\Sigma_i$ at $\alpha=\pi/2$, which we can parameterize with $(\theta,\phi)$,
and the twisted sector fields correspond to reductions of the 10d fields on these cycles.
In addition, since the internal space is compact, there are also $n-1$ dual 2-cycles $\tilde{\Sigma}_i$,
which we can parameterize with $(\alpha,\psi)$ (see Fig.~\ref{CyclesCartoon} for a cartoon sketch).
The orientifold 
acts on the cycles as 
\be
\label{orientifold_cycles}
\Omega I: \Sigma_i \rightarrow \Sigma_{n-i} \; , \;  \tilde{\Sigma}_i \rightarrow -\tilde{\Sigma}_{n-i} \,.
\ee
However for even orbifolds, as we saw, there are two possibilities for the projection in the
$k$-twisted sector.
This choice of ``discrete torsion" 
can be expressed in terms of the orientifold action on the middle cycles as
\be
\label{orientifold_middle_cycles}
\Omega I:
\Sigma_k \rightarrow 
\left\{
\begin{array}{rl}
\Sigma_k & \mbox{no VS} \\
-\Sigma_k & \mbox{VS}
\end{array}
\right.
\; , \;
\tilde{\Sigma}_k \rightarrow 
\left\{
\begin{array}{rl}
-\tilde{\Sigma}_k & \mbox{no VS} \\
\tilde{\Sigma}_k & \mbox{VS}
\end{array}
\right.\,.
\ee

\subsection{Charged mesons}

The global symmetries of the quiver theories should correspond to massless gauge fields in
the supergravity background. 
First, there are gauge fields associated to the isometry group of the internal space $SU(2)\times U(1)$.
The $SU(2)$ comes from the $S^2$ base of the Lens space $S^3/\mathbb{Z}_n$, and is dual to the 
R-symmetry, and the $U(1)$ comes from the $S^1$ fiber $\psi$, and is dual to the mesonic symmetry $U(1)_M$.
These symmetries are common to all the models with $n>2$.
For $n=2$ the isometry is enhanced to $SU(2)\times SU(2)$, in agreement with the enhanced mesonic symmetry.
The charged mesons correspond to KK states carrying momentum in $\psi$.
Since $\psi\sim \psi + 4\pi/n$ these states must carry an integer multiple of $n/2$ units of momentum,
{\em i.e.}, $k$ for the even orbifolds and $k+1/2$ for the odd orbifolds, 
in agreement with the mesonic charges found in the quiver theories.

\subsection{Baryons}

As in many other examples of AdS/CFT, the baryonic symmetries correspond to gauge fields
that are obtained by reducing higher-rank RR forms on finite cycles of the internal space,
and the baryons themselves correspond to D-branes wrapped on these cycles.
In our case the baryons are described by D2-branes wrapped on the 2-cycles $\tilde{\Sigma}_i$,
and the corresponding gauge field by the reduction of the 3-form $C_3$ on $\tilde{\Sigma}_i$.

The action of the orientifold on a non-middle cycle, combined with the action on $C_3$, leaves invariant only
the combination 
\be
\tilde{C}_{1,i} = \int_{\tilde{\Sigma}_i} C_3 + \int_{\tilde{\Sigma}_{n-i}} C_3\,,
\ee
and correspondingly only the state with a D2-brane on $\tilde{\Sigma}_i$ plus a D2-brane on $\tilde{\Sigma}_{n-i}$.
This state is dual to a di-baryon operator $B_i$ in the quiver field theory.
In particular there are $k-1$ of them in the even orbifolds and $k$ in the odd orbifold. 
The $AdS_6$ mass of such a state is given by
\be
\label{dibaryon_mass}
m_{B} &=& 2 \mu_2 e^{-\Phi} 
\int \hat{\Omega}(\alpha) \sqrt{g_{\alpha\alpha} g_{\psi\psi}} \, d\alpha \, d\psi  \nonumber\\
&=& \left(\frac{3^4 (8-N_f)}{(4\pi)^4}\right)^{1/3} \frac{Q_4^{3/4}}{n} \,.
\ee
For large $m_{B}\,L$ the dimension of the corresponding operator should be $\Delta = m_{B}\,L$.
Inserting the appropriate values of $Q_4$ and $L$, we find $\Delta = 3N$, in agreement with the dimension
of the di-baryon operators. 

In the even orbifold {\em without} vector structure there is an additional
gauge field from the middle cycle $\tilde{C}_{1,k} = \int_{\tilde{\Sigma}_k} C_3$,
and an additional charged state corresponding to a D2-brane on $\tilde{\Sigma}_k$,
which is dual to the Pfaffian operator $B_A$.
Since the D2-brane wraps only one cycle in this case its mass is half of (\ref{dibaryon_mass}),
and therefore $\Delta=3N/2$, in agreement with the dimension of $B_A$.

\subsection{Other branes}

Let us briefly consider additional brane wrappings and their interpretation in the dual field theories.

\subsubsection{Instantons}

The quiver theories possess an instantonic $U(1)_I$ symmetry for each gauge group factor.
The dual supergravity gauge fields include the
RR 1-form $C_1$ and the reductions of the RR 3-form $C_3$ on the vanishing 2-cycles $\Sigma_i$.
In particular, the 1-form is dual to the diagonal instanton symmetry.
The corresponding instantons are dual to the D0-brane and to D2-branes wrapping $\Sigma_i$,
namely to fractional D0-branes. The orientifold action on the non-middle cycles leaves only the combinations
\be
C_{1,i} = \int_{\Sigma_i} C_3 - \int_{\Sigma_{n-i}} C_3\,, 
\ee
and therefore the states with a D2-brane on $\Sigma_i$ and an anti-D2-brane on $\Sigma_{n-i}$.
For the even orbifold {\em with} vector structure there is an additional gauge field
$C_{1,k} = \int_{\Sigma_k} C_3$, and an additional fractional D0-brane described by a D2-brane on $\Sigma_k$.
In the orbifold without vector structure it is projected out.

In all, there are $k+1$ gauge fields associated to fractional D0-branes in the odd orbifold
and in the even orbifold with vector structure, and $k$ of them in the even orbifold without vector structure,
in agreement with the instantonic symmetries of the quiver theories.
As in the parent theory, the fractional D0-branes must have strings attached to them due
to the coupling to $F_0$, in agreement with the attachment of the Wilson lines to the instantons.

\subsubsection{Cosmic strings} 

A D4-brane wrapping a 2-cycle in the internal space corresponds to a membrane in $AdS_6$.
If it is localized in the radial direction it describes a membrane, namely a co-dimension 2 object,
or cosmic string, in the dual 5d field theory.
There are two types depending on whether the D4-brane wraps a vanishing cycle $\Sigma_i$
(together with its image $\Sigma_{n-i}$) or a dual cycle $\tilde{\Sigma}_i$ (together with its image).
The corresponding cosmic string sources a monodromy for the bulk gauge field
$\tilde{C}_{1,i}$ and $C_{1,i}$, respectively.
We can therefore refer to them as baryonic and instantonic cosmic strings, respectively.
In going around a baryonic (instantonic) cosmic string, the phase of a baryon (instanton)
described by a D2-brane wrapped on $\tilde{\Sigma}_i$ ($\Sigma_i$) goes through a full $2\pi$ rotation.
There is one more cosmic string corresponding to a D6-brane wrapped on the whole internal space.
This sources a monodromy for $C_1$, and is therefore associated to the diagonal instanton
dual to the D0-brane.

All of the cosmic strings are tensionless since gravity pulls them down to the origin of $AdS_6$.
This is as expected in the dual field theories, since the global symmetries are unbroken.
The broken phases are described by blowing up the original orbifold (see \cite{Klebanov:1999tb, Klebanov:2007us} for a general discussion, 
and  \cite{Klebanov:2007cx, Benishti:2010jn} for the relevance of the cosmic strings in the holographic realization of spontaneous symmetry breaking).
The dual backgrounds are asymptotically $AdS_6$, but they terminate at a radial position related
to the blow-up parameter. The tension of the baryonic cosmic strings is propotional to this parameter through the volume of the particular blown-up cycle which they wrap, which in turn corresponds in a precise sense to the particular baryonic $U(1)$ which is spontaneously broken. Note that the blow up corresponds to VEVs for the bi-fundamental fields, each of which breaks both a baryonic $U(1)$ symmetry and an instantonic $U(1)_I$ symmetry. Thus, we expect also the instantonic strings to acquire a non-zero tension.

\subsubsection{Domain walls}
\label{domainwalls}

A D6-brane wrapping a 2-cycle corresponds to a domain wall in $AdS_6$.
As for the cosmic strings, there are two types of domain walls.
A D6-brane wrapping one of the vanishing cycles $\Sigma_i$ (and its image)
corresponds to a ``fractional D4-brane" which changes 
one relative rank of the gauge groups.
This is a ``baryonic domain wall" in the sense that when the baryonic D2-brane 
wrapping $\tilde{\Sigma}_i$ crosses it a string is created between them.
This corresponds in the field theory to a Wilson line in the fundamental representation that must 
be added to the baryon to saturate the additional color index.
The number of different baryonic domain walls should equal the number of gauge groups in the quiver theory minus one.
There are $k-1$ ($k$) domain walls corresponding to the non-middle cycles in the even (odd) orbifolds,
and one more from the middle cycle in the even orbifolds with vector structure (since $\Sigma_k$ is odd in this case).
Indeed the quiver theories for the odd orbifolds and even orbifolds with vector structure have $k+1$ gauge groups,
and the quiver theories for the even orbifolds without vector structure have $k$ gauge groups.

The other type of domain wall corresponds to a D6-brane wrapping one of the dual cycles $\tilde{\Sigma}_i$.
This is an ``instantonic domain wall" that changes the bare CS level of one of the ($SU$) gauge groups by one. 
Now it is the D2-brane wrapping $\Sigma_i$ that picks up a string when it crosses the D6-brane wall. 
This corresponds to the additional Wilson line added to the dual instanton upon changing the CS level.
The number of different instantonic domain walls should equal the number of $SU$ groups in the quiver theory.
The counting is similar to the baryonic domain walls, except that the middle cycle only contributes
for the even orbifold {\em without} vector structure in this case.
Thus there are $k$ instantonic domain walls in the odd orbifolds and even orbifolds without vector structure,
and $k-1$ in the even orbifolds with vector structure,
in precise agreement with the number of $SU$ factors in the corresponding quiver theories.

Adding a D6-brane domain wall to the supergravity background sources an $F_2$ flux on the complementary 2-cycle.
In the models we considered $F_2=0$, and hence the ranks of gauge groups are equal and the bare CS terms vanish.

Another logical possibility for a domain wall would be to wrap a D8-brane on the entire $S^4/\mathbb{Z}_n$,
however this is unstable due to the orientifold projection. 
This is consistent with the fact that the $USp$ factors do not admit a bare CS interaction.

\subsubsection{(No) Baryon vertices}

A D4-brane wrapping the whole internal space, together with the required $2N$ strings, 
would correspond to a baryon vertex composed of $2N$ external fermions in the fundamental
representation of one of the gauge groups:
${\cal B} =  \epsilon_{a_1\cdots a_{2N}}\psi^{a_1}\cdots \psi^{a_{2N}}$.
Like the D8-brane, this is unstable due to the orientifold projection.
However it is worthwhile understanding how this instability manifests itself in the dual field theories.
The baryon vertex will generically decay into states corresponding to external mesons
and possibly local baryonic states.
Let us consider the different cases.

In the parent $USp(2N)$ theory the external fermions $\psi^a$
transform in the fundamental representation of $USp(2N)$ and are therefore pseudoreal.
The baryon vertex in this case can decay to $N$ external mesons of the form ${\cal M} = J_{ab}\psi^a\psi^b$,
since ${\cal B}\propto {\cal M}^N$
(a similar thing happens in the 4d ${\cal N}=4$ $USp(2N)$ theory \cite{Witten:1998xy}).

In the orbifolds theories that contain $USp(2N)$ factors (the odd orbifolds and the even orbifolds with vector structure)
we get basically the same result for a baryon vertex made of $USp$ external fermions.
In these theories one can also construct baryon vertices using external fermions
in the (complex) fundamental representation of one of the $SU(2N)$ factors.
These cannot decay purely into external mesons since one cannot form mesons purely out of $SU$ fundamentals.
The mesons must also include matter hypermultiplets, which means that the decay products include also di-baryons.
Consider for example the $\mathbb{Z}_3$ orbifold.
The dual field theory has $G=USp(2N)\times SU(2N)$, a bi-fundamental hypermultiplet $x$
and an $SU(2N)$ antisymmetric $A$.
The external meson in this case is given by 
${\cal M} = J_{ac}x^{a}_{\bar{b}}x^{c}_{\bar{d}}\psi^b\psi^d$.
The analogous relation in this case is ${\cal B} B \propto {\cal M}^N$, where $B=\mbox{det}(x)$ is the di-baryon operator.
Therefore, at least as far as charges are concerned,
the baryon vertex can decay to $N$ external mesons plus an anti-di-baryon.\footnote{Alternatively,
we could describe the external meson as ${\cal M} = \mbox{Tr}(\psi^2\bar{A}) = \psi^a\psi^b \bar{A}_{\bar{a}\bar{b}}$.
This would lead us to conclude that the baryon vertex can decay into $N$ external mesons plus a
Pfaffian baryon, which is just a change of basis for the baryons.}
This matches nicely with the supergravity description.
The $USp$ baryon vertex corresponds to a D4-brane wrapping $S^4/\mathbb{Z}_3$, and
the $SU$ baryon vertex corresponds to a D4-brane wrapped on $S^4/\mathbb{Z}_3$ with worldvolume
flux on the vanishing cycle $\Sigma_1$ (and its image $\Sigma_2$), 
corresponding to an additional D2-brane wrapped on $\tilde{\Sigma}_1$ (and its image $\tilde{\Sigma}_2$).

In the even orbifolds without vector structure there are only $SU$ baryon vertices.
As in the previous case, these can decay to $N$ external mesons plus some baryons.
For example in the $\mathbb{Z}_2$ case, namely $SU(2N)$ with two antisymmetrics $A,A'$,
we have the relation ${\cal B} \bar{B}_A \propto {\cal M}^N$, where now ${\cal M} = \mbox{Tr}(\psi^2 \bar{A})$.
Thus the baryon vertex ${\cal B}$ can decay into $N$ external mesons plus a Pfaffian baryon.
Note that unlike in the orbifolds with vector structure, one cannot get rid of the baryon.
There is no baryon vertex that decays solely into external mesons.
This fits nicely with the supergravity picture. In the orbifold without vector structure there
is trapped $B_2$ flux $b=1/2$ on the vanishing (middle) cycle $\Sigma_1$,
which implies that there is always a D2-brane wrapped on $\tilde{\Sigma}_1$ inside the wrapped D4-brane.
Turning on an integer woldvolume flux cannot remove it.

More generally the decay of a baryon vertex associated to any $SU(2N)$ factor
in the $SU(2N)^k$ quiver will involve all the bi-fundamental fields connecting it to one of the ends of the quiver,
as well as the antisymmetric field at that end.
The decay products will therefore include all the corresponding di-baryons $B_i$
as well as a Pfaffian baryon.

\subsection{Quark-antiquark potential}

As a simple application of the duality, let us compute the quark-anti-quark potential in the 5d quiver fixed point theories. 
Using the standard prescription, 
we consider a string with worldsheet coordinates $(t,x)$, and take $u=u(x)$
with $0\leq x \leq \ell$.
The standard minimal area computation yields
\begin{equation}
S=-\frac{2\,T_{F1}\,I_2\,I_{-2}}{\ell}\,\frac{9}{4}\,\sqrt{Q_4}\,\hat{\Omega}^2(\alpha) \,,
\end{equation}
where
\begin{equation}
I_p \equiv \int_1^{\infty}\,dy\,\frac{y^p}{\sqrt{y^4-1}} \,.
\end{equation}
As usual $I_2$ diverges, and must be renormalized by subtracting the contribution of the 
free quark-antiquark pair. 
This amounts to replacing 
\begin{equation}
I_{2}\rightarrow -1+ \int_1^{\infty}\,dy\,(\frac{y^2}{\sqrt{y^4-1}}-1) \,.
\end{equation}
Since $\hat{\Omega}(\alpha)\propto \sin^{-1/6}\alpha$, the action is minimized at $\alpha=\pi/2$. 
The resulting potential is given by
\begin{equation}
V(\ell)=-S = -\frac{6\,\sqrt{2}\,\pi^2\,\Gamma\Big(\frac{3}{4}\Big)}{\Gamma\Big(\frac{1}{4}\Big)^3}\,
\sqrt{\frac{N}{8-N_f}}\,\frac{1}{\ell}\,\sim\,-2.15\,\sqrt{\frac{N}{8-N_f}}\,\frac{1}{\ell} \,.
\end{equation}

%%%%%%%%%%%%%%%%%%%%%%%%%%%%%%%%%%%%%%%%%%%%%%%%%%%%%%%%%%%%

\section{Conclusions}

In this paper we have initiated a study of supergravity duals of 5d supersymmetric fixed points associated 
to quiver gauge theories.
We identified three classes of theories corresponding to D4-branes in orbifolds of Type I' string theory,
and related them to warped products of $AdS_6$ and $S^4/\mathbb{Z}_n$ in massive Type IIA supergravity.
We also analyzed the gauge symmetries, charges and branes in the supergravity backgrounds, and found a complete agreement with the quiver gauge theories.

These new examples of AdS/CFT dual pairs suggest several directions for further exploration.
First, a detailed analysis of the Kaluza-Klein spectrum should be performed and compared with 
the spectrum of chiral primary operators in the gauge theories. 

There are also a number of natural generalizations that one can study.
As we alluded to in the discussion of domain walls, turning on a RR 2-form flux on a 2-cycle
of the internal space should correspond, depending on the 2-cycle, either to changing
the relative ranks of the gauge groups or to adding a bare CS coupling to $SU$ factors.
However the backreaction of such fluxes needs to be analyzed.
It would also be interesting 
to construct the supergravity 
duals of RG flows triggered by VEVs of baryonic operators along the lines of \cite{Klebanov:2007us}.

One of the outstanding questions concerns the singularity at $\alpha=0$.
On the one hand, this singularity is responsible for the enhancement of the global symmetry at the origin of 
the Coulomb branch to $E_{N_f+1}$,
a feature common to all the orbifold theories which follows from the D4-D8-O8 description. 
As this symmetry should be manifest along the Higgs branch, it would be very interesting to consider in more detail the 
Higgs branches of the theories.
On the other hand one may wonder whether the singularity can be resolved, 
in view of the conjecture of \cite{Aharony:2010af} 
that massive Type IIA supergravity cannot be strongly coupled. 
Nevertheless, the backgrounds discussed here are more complicated since in addition to strong coupling, 
the curvature is large.

Another central issue that needs to be understood better is the fate 
of the singularities on the moduli space of the quiver theories
\cite{Intriligator:1997pq}.
The $AdS_6\times S^4/\mathbb{Z}_n$ backgrounds provide a strong indication
that fixed points exist in the corresponding quiver theories, but
it would be interesting to have a deeper understanding of the underlying field theory mechanism for these fixed points.
A key observation is that instanton particles become massless at the singular points of the Coulomb branch.
Taking these states into account may remove the singularities and lead to well defined quiver fixed point theories.
The 5-brane web constructions, when available, suggest a continuation past infinite coupling by S-duality,
whereby the massless instantons are exchanged for ordinary massless W-bosons. It would certainly be 
interesting to study this more systematically for any orbifold and any rank.
In particular, the 5-brane webs for odd orbifolds and even orbifolds without vector structure are not known,
since the resolution of the O7-plane with a stuck NS5-brane is not understood.
It is thus of great interest to clarify whether such a quantum resolution exists and, if so, what it is.

Finally, we wish to point out that even among the examples of 5d fixed points classified in \cite{Seiberg:1996bd,Morrison:1996xf,Intriligator:1997pq}, most of them still lack an AdS/CFT description. In particular, those corresponding to $SU$ gauge groups should admit a tunable CS coefficient. It is natural to wonder what the gravity duals for these might be.

\section*{Acknowledgements}

We are grateful to Andreas Brandhuber, Amihay Hanany and Dario Martelli for useful conversations. D.R-G thanks the Universidade de Santiago de Compostela, Witts University and the Newton Institute for Mathematical Sciences at Cambridge for warm hospitality while this work was in progress. D.R-G. is supported by the Aly Kaufman fellowship. He also acknowledges partial support from the Israel Science Foundation under grant no 
392/09 and from the Spanish Ministry of Science through the research grant FPA2009-07122 and Spanish Consolider-Ingenio 2010 Programme CPAN (CSD2007-00042). O.B. is supported in part by the Israel Science Foundation under grant no. 392/09,
and the US-Israel Binational Science Foundation under grant no. 2008-072.

\end{document}